\documentclass[aps,prl,preprint,tightenlines,superscriptaddress,showpacs,byrevtex]{revtex4}
\usepackage{graphicx,color} % Include figure files
\usepackage{dcolumn}  % Align table columns on decimal point

\graphicspath{{ps}}

\renewcommand{\arraystretch}{1.1}

\def\ra{\!\rightarrow\!}
\def\Taupipi0{\tau^{-}\!\ra\pi^{-}\pi^{0}\,\nu_{\tau}}
\def\Tauhpi0{\tau^{-}\!\ra h^{-}\pi^{0}\,\nu_{\tau}}
\def\Pipi0{\pi^{-}\pi^{0}}
\def\2piamu{a_{\mu}^{{\rm had},2\pi}}

\def\MassSQ{M^{2}_{\pi\pi^{0}}}
\def\MassSQobs{M^{2}_{\rm obs}}
\def\MassSQgen{M^{2}_{\rm gen}}

\def\GeVCC{{\rm GeV}/c^{2}}
\def\GeVcc2{({\rm GeV}/c^{2})^{2}}

\definecolor{mycolor}{gray}{0.8}

\begin{document}

\vspace*{-3\baselineskip}
\resizebox{!}{3cm}{\includegraphics{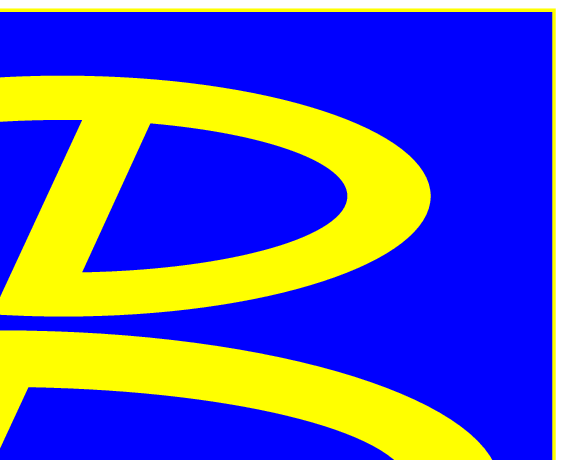}}

\preprint{\vbox{ \hbox{   }
	 \hbox{BELLE-CONF-0508}
\hbox{EPS05-480}
%\hbox{ }
%\hbox{Version 4.0 }
%                 \hbox{EPS05: Paper-143}
%                 \hbox{hep-ex nnnn, if availabl}
}}

\title{ \quad\\[0.5cm] 
A High Statistics Study of the Decay $\Taupipi0$   }

%%%% insert the authorlist here. BEFORE the abstract !!!
%%% Paper:    
%%% Journal:  Summer 2005 conference papers
%%% Contacts: 
%%% Non-responding authors or those who said NO are commented out.
%%% ====================================================================
%%% Click the RELOAD button on your web browser to see the updated file.
%%% ====================================================================
%%% Use \input{author} to insert this material into your latex file.
%%%%% Force institutions to appear in alphabetical order when typeset.
\affiliation{Aomori University, Aomori}
\affiliation{Budker Institute of Nuclear Physics, Novosibirsk}
\affiliation{Chiba University, Chiba}
\affiliation{Chonnam National University, Kwangju}
\affiliation{University of Cincinnati, Cincinnati, Ohio 45221}
\affiliation{University of Frankfurt, Frankfurt}
\affiliation{Gyeongsang National University, Chinju}
\affiliation{University of Hawaii, Honolulu, Hawaii 96822}
\affiliation{High Energy Accelerator Research Organization (KEK), Tsukuba}
\affiliation{Hiroshima Institute of Technology, Hiroshima}
\affiliation{Institute of High Energy Physics, Chinese Academy of Sciences, Beijing}
\affiliation{Institute of High Energy Physics, Protvino}
\affiliation{Institute of High Energy Physics, Vienna}
\affiliation{Institute for Theoretical and Experimental Physics, Moscow}
\affiliation{J. Stefan Institute, Ljubljana}
\affiliation{Kanagawa University, Yokohama}
\affiliation{Korea University, Seoul}
\affiliation{Kyoto University, Kyoto}
\affiliation{Kyungpook National University, Taegu}
\affiliation{Swiss Federal Institute of Technology of Lausanne, EPFL, Lausanne}
\affiliation{University of Ljubljana, Ljubljana}
\affiliation{University of Maribor, Maribor}
\affiliation{University of Melbourne, Victoria}
\affiliation{Nagoya University, Nagoya}
\affiliation{Nara Women's University, Nara}
\affiliation{National Central University, Chung-li}
\affiliation{National Kaohsiung Normal University, Kaohsiung}
\affiliation{National United University, Miao Li}
\affiliation{Department of Physics, National Taiwan University, Taipei}
\affiliation{H. Niewodniczanski Institute of Nuclear Physics, Krakow}
\affiliation{Nippon Dental University, Niigata}
\affiliation{Niigata University, Niigata}
\affiliation{Nova Gorica Polytechnic, Nova Gorica}
\affiliation{Osaka City University, Osaka}
\affiliation{Osaka University, Osaka}
\affiliation{Panjab University, Chandigarh}
\affiliation{Peking University, Beijing}
\affiliation{Princeton University, Princeton, New Jersey 08544}
\affiliation{RIKEN BNL Research Center, Upton, New York 11973}
\affiliation{Saga University, Saga}
\affiliation{University of Science and Technology of China, Hefei}
\affiliation{Seoul National University, Seoul}
\affiliation{Shinshu University, Nagano}
\affiliation{Sungkyunkwan University, Suwon}
\affiliation{University of Sydney, Sydney NSW}
\affiliation{Tata Institute of Fundamental Research, Bombay}
\affiliation{Toho University, Funabashi}
\affiliation{Tohoku Gakuin University, Tagajo}
\affiliation{Tohoku University, Sendai}
\affiliation{Department of Physics, University of Tokyo, Tokyo}
\affiliation{Tokyo Institute of Technology, Tokyo}
\affiliation{Tokyo Metropolitan University, Tokyo}
\affiliation{Tokyo University of Agriculture and Technology, Tokyo}
\affiliation{Toyama National College of Maritime Technology, Toyama}
\affiliation{University of Tsukuba, Tsukuba}
\affiliation{Utkal University, Bhubaneswer}
\affiliation{Virginia Polytechnic Institute and State University, Blacksburg, Virginia 24061}
\affiliation{Yonsei University, Seoul}
  \author{K.~Abe}\affiliation{High Energy Accelerator Research Organization (KEK), Tsukuba} % KEK
  \author{K.~Abe}\affiliation{Tohoku Gakuin University, Tagajo} % TohokuGakuin
  \author{I.~Adachi}\affiliation{High Energy Accelerator Research Organization (KEK), Tsukuba} % KEK
  \author{H.~Aihara}\affiliation{Department of Physics, University of Tokyo, Tokyo} % Tokyo
  \author{K.~Aoki}\affiliation{Nagoya University, Nagoya} % Nagoya
  \author{K.~Arinstein}\affiliation{Budker Institute of Nuclear Physics, Novosibirsk} % BINP
  \author{Y.~Asano}\affiliation{University of Tsukuba, Tsukuba} % Tsukuba
  \author{T.~Aso}\affiliation{Toyama National College of Maritime Technology, Toyama} % Toyama
  \author{V.~Aulchenko}\affiliation{Budker Institute of Nuclear Physics, Novosibirsk} % BINP
  \author{T.~Aushev}\affiliation{Institute for Theoretical and Experimental Physics, Moscow} % ITEP
% ...
% Lausanne
\collaboration{The Belle Collaboration}

%\collaboration{Belle Collaboration}
%\noaffiliation

\begin{abstract}
We report a measurement of the branching fraction 
for $\Taupipi0$ and the invariant mass spectrum 
of the resulting $\pi^{-}\pi^{0}$ system 
using $72.2~{\rm fb}^{-1}$ of data recorded 
by the Belle detector at the KEKB $e^+e^-$ collider.
The branching fraction obtained is 
$(25.15\pm 0.04 \pm 0.31)\% $,
where the first error is statistical and the second is systematic.
The unfolded $\pi^{-}\pi^{0}$ mass spectrum
is used to determine resonance parameters for the 
$\rho(770)$, $\rho'(1450)$, and $\rho''(1700)$ mesons.
We also use this spectrum to estimate the hadronic contribution 
to the anomalous magnetic moment of the muon.
\end{abstract}

%\pacs{13.65.+i, 13.25.Gv, 14.40.Gx}
%
% 14.60.Fg  :Tau Leptons
% 13.35.Dx  : Tau Decays
% 13.40.Gp   : Electromagnetic Form Factor
%\date{ 01/Dec/2005      }
\maketitle

{\renewcommand{\thefootnote}{\fnsymbol{footnote}}}
\setcounter{footnote}{0}
%$\mathcal{B}$, $\mathrm{B}$,$\mathit{B}$, $\mathtt{B}$,

\section{Introduction}

Among the decay channels of the $\tau$ lepton, $\Taupipi0$
has the largest branching fraction. The decay is dominated by
intermediate resonances and thus provides information on the 
properties of the $\rho(770)$, $\rho^{\prime}(1450)$, and 
$\rho^{\prime\prime}(1700)$ mesons and their interference.
Since leptons do not participate in the strong interaction,
hadronic $\tau$ decays provide a clean environment for  
studying the dynamics of hadronic states
in an interesting energy range dominated by resonances.

Under the Conserved Vector Current~(CVC) theorem,
the  $\pi^{-}\pi^{0}$ mass spectrum in this range
%is of great importance for improving 
can be used to improve the theoretical error 
on the anomalous magnetic moment of the muon 
$a_{\mu}=(g_{\mu}-2)/2$. A recent review of the 
calculations of $a_{\mu}$ is given in Ref.~\cite{CM}.   
It is known that the theoretical error on $a_{\mu}$ is dominated by 
the (leading-order) hadronic contribution 
$a_{\mu}^{\rm had,LO}$, given by the hadronic vacuum polarization.
This contribution cannot be evaluated within the framework of
perturbative QCD; however, it can be evaluated from a
measurement of the cross section for $e^+e^-$ annihilation 
to hadrons~\cite{DEHZ,HMNT}. Alternatively, 
CVC 
%and isospin invariance
relates the properties of 
the $\pi^{+}\pi^{-}$ system produced 
in $e^+e^-\rightarrow\pi^+\pi^{-}$ to those of the 
$\pi^{-}\pi^{0}$ system produced in $\Taupipi0$ decay; 
thus, using CVC and correcting for 
%Based on CVC and corrected for 
isospin-violating effects, 
$\tau$ data have also been used to obtain a
 more precise
 prediction for $a^{\rm had,LO}_{\mu}$~\cite{DH98,DEHZ}.

Recently,  data on $e^+e^-\!\rightarrow\!\pi^+\pi^-$  
has become available from the CMD-2, KLOE, 
and SND experiments~\cite{CMD03,CMD2002,KLOE2005,SND2005}. 
Data on $\tau$ decays is available from the
ALEPH~\cite{ALEPH97,DATAU02}, CLEO~\cite{CLEO2000}, 
and OPAL~\cite{OPAL1999} experiments. The most recent evaluation 
of the hadronic contribution to $a_{\mu}$ using
$e^+e^-$ data gives~\cite{DAV2005}
$ a^{\rm exp}_{\mu} - a^{\rm th}_{\mu} = (25.2\pm 9.2  )\times 10^{-10}$,
 while that using the  $\tau$ lepton data 
where applicable gives
$ a^{\rm exp}_{\mu} - a^{\rm th}_{\mu} = (9.4\pm 10.5  )\times 10^{-10}$.
%if  $\tau$ data are used for the relevant low-energy region
%for $2\pi$ and $4\pi$ modes.
%the most recent result using $\tau$ data is~\cite{DEHZ}
%In using the $\tau$ data, 
The experimental value $a_\mu^{\rm exp}$
%the predictions are compared to  the  world average of 
%the $a_{\mu}$ measurements, which 
is dominated by the BNL E821 measurement~\cite{BNL2004}
%$a^{\rm E821}_{\mu} = 
$( 11\ 659\ 208 \pm 5.8)\times 10^{-10}$.
%For the $\tau$-based result, isospin-breaking corrections
%are taken into account according to Ref.~\cite{ISB2001}. 
These differences correspond to 2.7 and 0.9 standard 
deviations, respectively, and thus 
there is a significant difference 
between the $e^+e^-$-based and 
$\tau$-based predictions. 
To clarify the situation, 
more data for $e^+e^-\!\rightarrow\!\pi^-\pi^+$ and for
$\tau^-\!\rightarrow\!\pi^-\pi^0\nu_{\tau}$
 decays are needed.
%and better information  on the
%$e^+e^-$ hadronic cross section as well as  on the
%hadronic mass spectra in $\tau$  decays are needed. 
In this paper we present a high-statistics measurement 
of  the $\pi^{-}\pi^{0}$ mass spectrum produced in
$\Taupipi0$ decays~\cite{CC} using 
%a large data sample 
data collected by the Belle experiment at the 
KEKB $e^+e^-$ collider operating at a 
center-of-mass (CM) energy of~10.6~GeV. 
The data sample is about 50 times larger 
than those of previous experiments.

\section{Basic formulae}

The differential decay rate for
$\Taupipi0$ can be expressed as
\begin{eqnarray}
\frac{1}{\Gamma}\frac{d\Gamma}{ds}
(\Taupipi0) =
\frac { 6\pi |V_{ud}|^{2} S_{EW} } {m_{\tau}^{2}} 
\left(
\frac{\mathcal{B}_{e}} { \mathcal{B}_{\pi\pi} } \right)
     \left( 1 - \frac{s}{m_{\tau}^{2}} \right)^{2}
     \left( 1 + \frac{2s}{m_{\tau}^{2}} \right)
\,v_{-}(s),
\label{eq:tauspec} 
\end{eqnarray}
where $s$ is the invariant-mass-squared of the $\pi^{-}\pi^{0}$ system,
$v_{-}(s)$ is the vector spectral function
characterizing the $\pi^{-}\pi^{0}$ system,   
$|V_{ud}|$ denotes the CKM mixing matrix element, and 
$S_{EW}$ accounts for electroweak radiative corrections. 
$\mathcal{B}_{e}$ 
and $\mathcal{B}_{\pi\pi}$ 
are the branching fractions for 
$\tau^{-}\rightarrow e^{-}\nu_{\tau}\bar{\nu}_{e}$ and
$\Taupipi0$, respectively. 

The corresponding $\pi^{+}\pi^{-}$spectral function $v_{0}(s)$ can
be obtained from the $e^+e^-\rightarrow \pi^+\pi^-$ cross section:
\begin{eqnarray}
\sigma(e^+e^-\rightarrow\pi^+\pi^-)  = 
\frac{4\pi\alpha_{0}^{2}}{s}\,v_{0}(s),
\label{eq:eepipi}
\end{eqnarray}
where $s$ is the $e^+e^-$ CM energy squared and
$\alpha_{0}$ is the fine-structure constant  at $s=0$.
Up to isospin-violating effects, CVC allows one to 
relate the spectral function from $\tau$ decays
to the isovector part of the $e^+e^-$ spectral function
\cite{Weakf}:
\begin{eqnarray}
v_{-}(s) & = & v_{0}^{I=1}(s)\,.
\label{eq:cvc}
\end{eqnarray}

The mass spectrum of the two-pion system is typically expressed 
in terms of pion form factors; these are useful for comparing  
resonance shapes in the charged and neutral two-pion systems. 
The spectral function $v_{j}(s)\,(j=-,0)$ is related to the 
form factor $F^{j}_{\pi}(s)$ via
\begin{equation}
v_{j}(s) = \frac{\beta_{j}^{3}(s)}{12\pi}|F_{\pi}^{j}(s)|^{2},
\label{eq:pionform}
\end{equation} 
where $\beta_{-}(s)\, (\beta_{0}(s))$ 
is the pion velocity in the $\pi^-\pi^{0}$\, ($\pi^{+}\pi^{-}$) rest-system.
The velocities $\beta_{j}(s)$ are explicitly given by\
$\beta^2_{-}(s) = 
\left[ 1 - (m_{\pi^{-}} - m_{\pi^{0}})^2/s\right]
\left[ 1 - (m_{\pi^{-}} + m_{\pi^{0}})^2/s\right]$\, 
and\, $\beta^2_{0}(s) = 
\left[ 1 - 4m_{\pi^{-}}^2 /s\right]$ \,.
%The mass difference of $\pi^{-}$ and $\pi^{0}$ 
%results in a phase space difference proportional to
%$\beta^{3}_{0}/\beta^{3}_{-}$.

%\begin{equation}
% \beta_{-}(s) = \sqrt{ \left( 
%                           1 - \frac{(m_{\pi^{-}} - m_{\pi^{0}})^{2}}{s}
%                         \right)
%                         \left( 
%                           1 - \frac{(m_{\pi^{-}} + m_{\pi^{0}})^{2}}{s}
%                         \right)  }\,.
%\end{equation}
%The mass difference between $\pi^{-}$ and $\pi^{0}$ 
%results in a phase-space difference proportional to
%$\beta^{3}_{\pi^{-}\pi^{0}}/\beta^{3}_{\pi^{-}\pi^{+}}$.

%\subsection{Anomalous muon magnetic moment} 

The leading-order hadronic contribution to the muon anomalous 
magnetic moment ($a_{\mu}^{\rm had,LO}$) is related to the
$e^+e^-$ annihilation cross section via the dispersion integral
\begin{eqnarray}
a^{\rm had,LO}_{\mu}
&=& \left( \frac{\alpha_{0} m_{\mu}}{3\pi} \right)^{2}
\int_{4m_{\pi}^{2}}^{\infty} \frac{R(s)}{s^{2}} \hat{K}(s)\,ds\,,
\label{eq:amu}
\end{eqnarray}  
where $s$ is the invariant-mass-squared of the two-pion system, and 
$R(s) = \sigma(e^+e^-\!\rightarrow\!{\rm hadrons})/(4\pi\alpha_{0}^{2}/3s)$.
The kernel $\hat{K}(s)$  is a smooth function increasing from 0.63 at
the threshold $s=4m_{\pi}^{2}$ to unity at $s\!=\!\infty$~\cite{KERNEL}.
Due to the $1/s^{2}$ dependence, hadronic final states at low energy 
%is strongly enhanced in 
dominate the contribution to $a_{\mu}^{\rm had,LO}$; 
in fact about 70\% of $a_{\mu}^{\rm had,LO}$ is due to the two-pion 
%intermediate 
state having $4m_{\pi}^{2}\le s\le 0.8$  $\GeVcc2$.
Consequently, the $2\pi$ spectral function in 
%the $e^+e^-$ and 
$\tau$ data is 
%vitally important 
useful to obtain predictions for $a_\mu^{\rm had,LO}$.
Using Eqs.~(\ref{eq:eepipi}) and (\ref{eq:cvc})
to evaluate (\ref{eq:amu}) we obtain
%The contribution from the $2\pi$ intermediate state 
%$a^{\pi\pi}_{\mu}$ can be obtained via 
\begin{eqnarray}
a^{\pi\pi}_{\mu}
&=& \left( \frac{\alpha_{0} m_{\mu}}{3\pi} \right)^{2}
%   \int_{4m_{\pi}^{2}}^{m^2_\tau}
\int_{4m_{\pi}^{2}}^{m_{\tau}^{2}}
\frac{3\, v_{-}(s)}{  s^{2}} \hat{K}(s)\,ds\,
+ ... \,,
\label{eq:amu2pi}
\end{eqnarray}  
%using Eqs.~(\ref{eq:eepipi}) and (\ref{eq:cvc}) above.
%$v_{0}(s)$ defined above.
\noindent
where ''$...$'' 
indicates the integral above the $m_{\tau}^{2}$ region.
To determine $v_-(s)$, 
%the two-pion spectral function in $\Taupipi0$, 
one must measure both the branching fraction for
$\Taupipi0$ and the $\pi^-\pi^0$ mass spectrum $(1/N)(dN/ds)$. 
Here we report new measurements for both of these.

%===========================================%
%   section 2 : Data and event selection    %
%===========================================%
\section{Data Sample and Selection Criteria }

The data sample used was collected by the Belle detector at 
the KEKB energy-asymmetric $e^{+}e^{-}$ collider~\cite{KEKB}. 
It is based on an integrated luminosity of $72.2~{\rm fb}^{-1}$ 
recorded at a CM energy of 10.6~GeV.
The Belle detector is a large-solid-angle magnetic spectrometer
consisting of a three-layer silicon-vertex detector (SVD), a 50-layer 
central drift chamber (CDC) for charged particle tracking
and specific ionization measurement~($dE/dx$),
an array of aerogel threshold Cerenkov counters~(ACC),
a barrel-like arrangement of time-of-flight scintillation counters~(TOF),
and an electromagnetic calorimeter~(ECL) comprised of CsI(Tl) crystals
located inside a superconducting solenoid coil that provides a 
1.5 T magnetic field. An iron flux-return located outside of 
the coil is instrumented
to identify muons and to detect $K_{L}^{0}$ mesons (KLM).
The detector is described in detail elsewhere~\cite{Belle}.

%In order to achieve  less systematics, 
%several efforts have been made in the analysis. Below, we first summarize
%the important feature of this analysis before describing its details:
%$\bullet$ To make a measurement independent, it is important to measure
% the mass spectrum as well as  the branching fraction.
% For this purpose, 
% event selection has been carried out by  two-steps:
%(i) selection of the generic $\tau$-pair event and (ii)  selection
%of the decay $\Taupipi0$.
%The $\tau$-pair event, selected in the first step, can be used for 
%the normalization in  the branching-fraction measurement. 
%Several control samples have been used to check 
%the normalization of the background Monte Carlo(MC) simulations.
%These MCs are used to estimate the remaining background in the final
%$\Taupipi0$ 
%sample, although the fraction is small ($2.3\%$).
%$\bullet$  Understanding  the quality of the $\gamma/\pi^{0}$ detection
%is the most
%important issue in this analysis. In addition to the calibration of the
%energy scale using the data, the $\pi^{0}$ mass resolution has been
%parameterized as a function the energy and the polar angle, so that
%the MC reproduce the data correctly. 
%$\bullet$ The acceptance and the bin-by-bin migration effects 
%of the mass spectrum are corrected
% by the singular-value-decomposition method, originally developed
%by the ALEPH collaboration~\cite{SVD}.

To study backgrounds and determine selection criteria,
%the detection efficiency, 
%and remaining backgrounds, 
we perform Monte Carlo~(MC) simulation studies 
%are performed 
for various processes. The KORALB/TAUOLA
program~\cite{TAUOLA,TAUOLA2004} is used for $\tau^+\tau^-$-pair
 generation,
the QQ generator~\cite{QQ} for ${\bar B B}$ and ${\bar q q}$ continuum
processes, the BHLUMI~\cite{BHLUMI} program for radiative Bhabha events,
the KKMC~\cite{KKMC} program for radiative $\mu^+\mu^-$-pair
 events, and the
AAFHB~\cite{AAFHB} program for two-photon processes.
The BHLUMI and KKMC programs include higher-order radiative 
corrections and are among the most accurate programs available. 
The detector response is simulated by a GEANT-based program~\cite{GEANT}. 
In order to realistically simulate beam-induced background, 
detector hits taken from randomly-triggered data are added 
to wire hits in the CDC and to energy deposits in the ECL.

%===== event selection (tau pair) =====%
\subsection{$\tau^+\tau^-$ pair selection}

The event selection consists of two steps. 
Initially, a sample of generic
$e^+e^-\rightarrow \tau^+\tau^-(\gamma)$ events are
selected with relatively loose criteria. 
From this sample
$\Taupipi0$ decays are then selected.
The number of  generic $\tau^+\tau^-$ events is used to determine
the $\Taupipi0$ branching fraction.

Generic $\tau^+\tau^-$ events are selected by requiring that
the number of charged tracks in an event  
be two or four with zero net charge; 
that each track have a momentum
transverse to the beam axis ($p_T$)
of greater than 0.1~GeV/$c$; and that
each track extrapolate to the interaction point (IP) 
within ${\pm1}$ cm transversely 
and within ${\pm5}$ cm along the beam direction.
To suppress background from Bhabha and $\mu^+\mu^-$ events, 
the reconstructed CM energies and
the sum of the momenta of 
the two leading tracks are required to be less 
than 9.0 GeV/$c$. The maximum $p_T$ among the tracks is 
required to be greater than 0.5~GeV/$c$.
%        in order to satisfy the trigger condition.
%- event vertex -%
Beam-related background is rejected by requiring that the position 
of the reconstructed event vertex be less than 0.5~cm from the IP
in the transverse direction and less than 2.5~cm from the IP
along the beam direction.
%The opening angle of two particles is required to be greater than 
%90^{\circ}.
The polar angle of the leading particle with respect to the
beam axis ($\theta^*$) in the CM frame is required
to be in the fiducial region of the detector:
${35^{\circ} < \theta^* < 145^{\circ} }$. 

\begin{figure}[t]
%   \begin{minipage}[t]{7.7cm}
\rotatebox{0}{\includegraphics*[width=0.410\textwidth,clip]
%{./Fig/tautau_mmvstheta.eps}} 
 {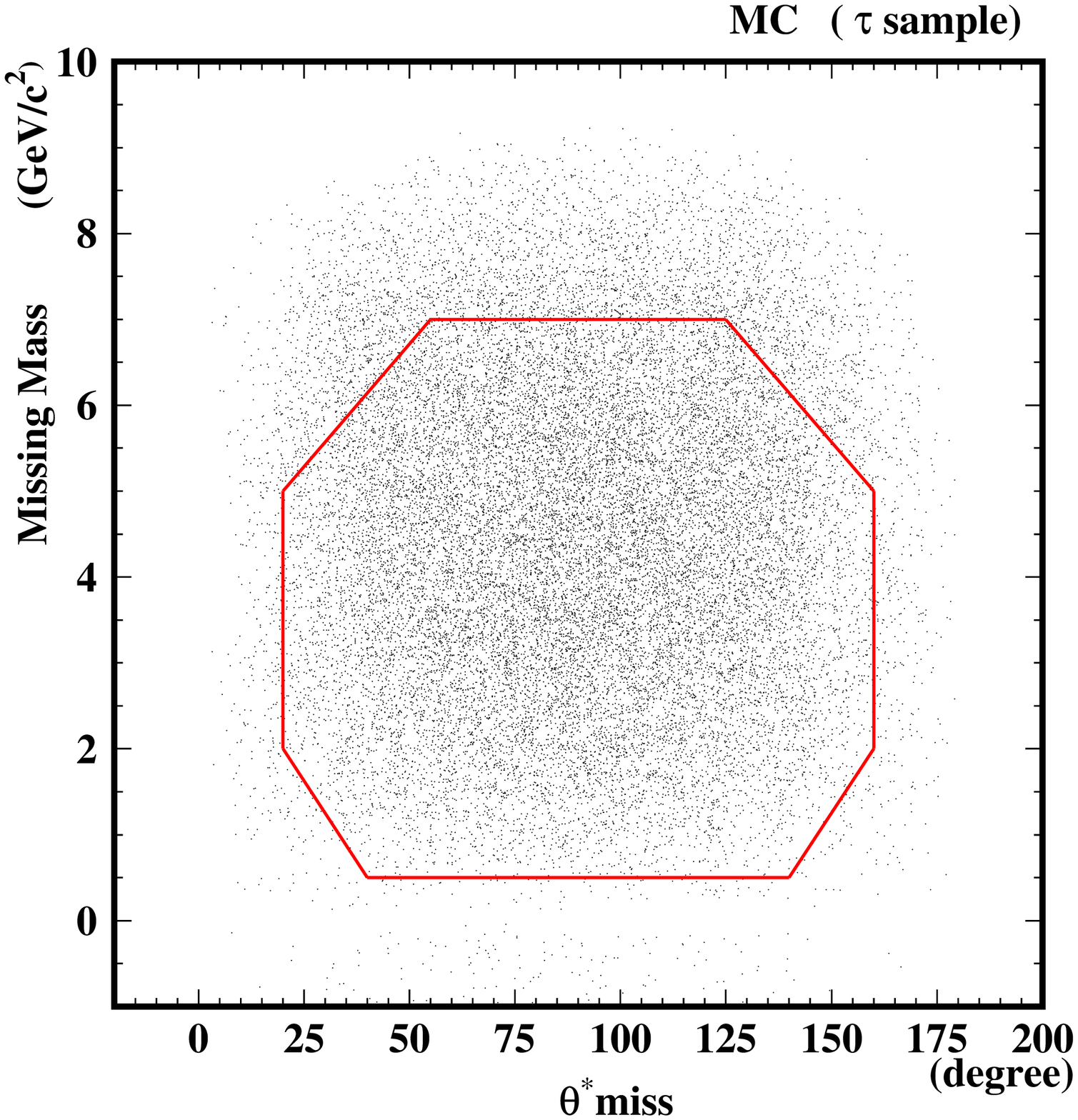}} 
%%%%
%
%%%%%
\rotatebox{0}{\includegraphics*[width=0.45\textwidth,clip]
%{./Fig/dat_mmvsthmiscm.eps}} 
%  
  {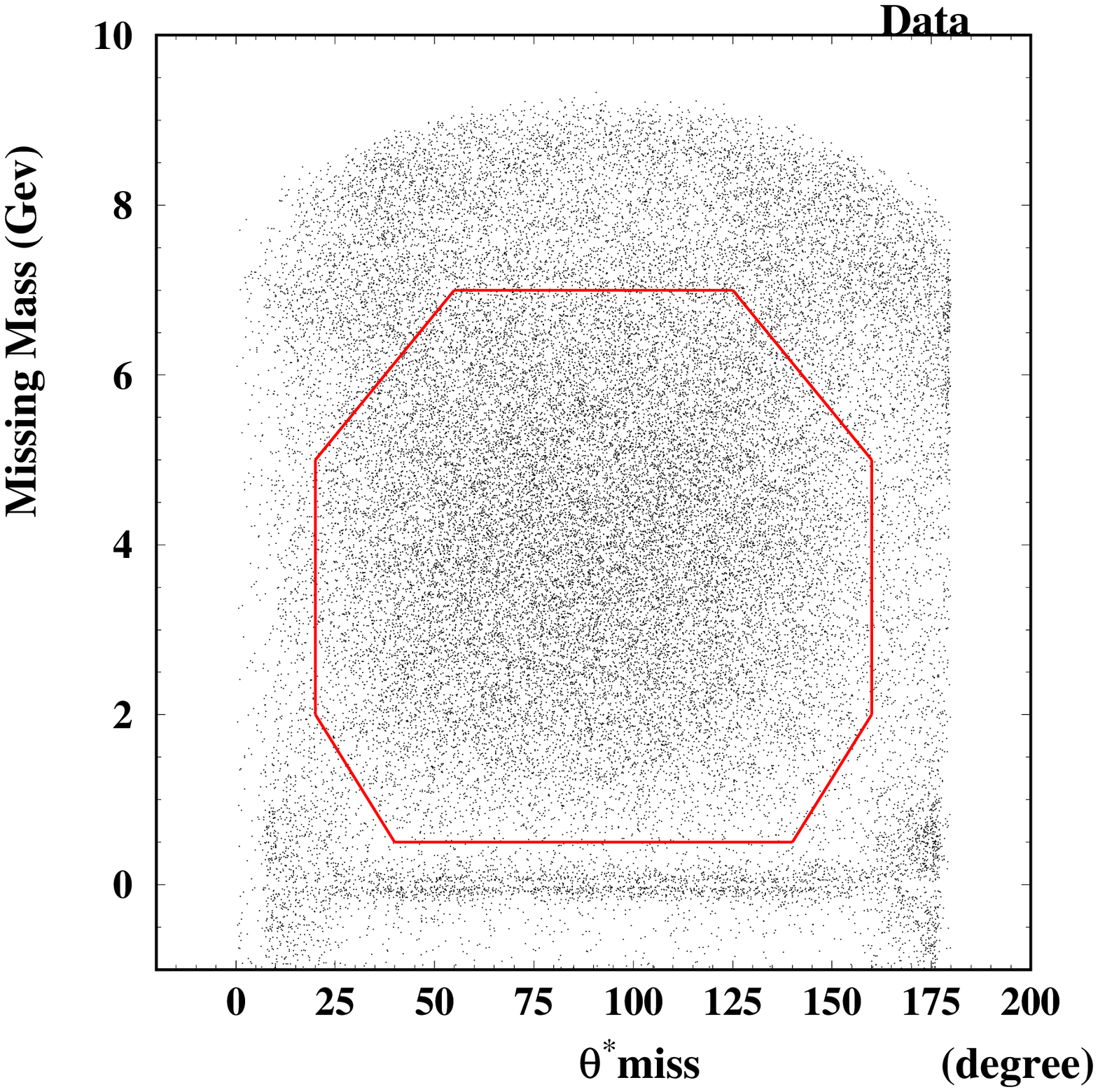}} 
%
%   }
\caption
{Missing mass ($MM$) versus  the polar 
angle for the direction 
of the missing momentum ($\theta^{*}_{\rm miss}$).
The left plot shows MC 
$e^+e^-\rightarrow \tau^+\tau^-$ events, and the right plot 
shows the data. Events inside the octagonal region are selected 
as $\tau^+\tau^-$-pair candidates.}
\label{data_mm}
\end{figure}

To reduce remaining background from Bhabha, $\mu^+\mu^-\gamma$,
and two-photon events,
a cut is applied in the two-dimensional plane of the
missing-mass $MM$ and the direction of missing momentum 
in CM ${\theta^{*}_{\rm miss}}$, where
$MM$ is evaluated from the four-momenta
of the measured tracks and  photons:
${(MM)^{2} = (P_{\rm ini} - P_{\rm tracks} 
- P_{\gamma s})^{2} }$. In this expression $P_{\rm ini}$
is the four-momentum of the initial $e^+e^-$ system.
Each photon (reconstructed from clusters in the calorimeter) must 
be separated 
from the nearest track projection by at least 20~cm
and have an energy greater than 0.05~GeV in the 
central part 
($-0.63\le\cos\theta< 0.85$),\, and 0.1~GeV in the endcap 
part ($-0.90\le\cos\theta< -0.62$ and $0.85\le\cos\theta<0.95$).
Photons measured at the 
detector edge are rejected. A scatterplot of 
$MM$ vs.\ $\theta_{\rm miss}$ 
for data is shown in Fig.~\ref{data_mm}.
In this plot, events at $MM\approx 0$ are due to radiative Bhabha
events and $e^+e^-\rightarrow \mu^+\mu^-(\gamma)$, 
while events in the high-$MM$ region are from two-photon 
processes.  
Events within the octagonal region
are selected as $\tau^+\tau^-$candidates.

%-------------------- missing mass ----------------------%
%--- Data ---
%----------------------------------------------------------%

Candidate events are divided into two hemispheres in the 
CM frame with respect to the highest momentum particle, and 
the remaining background from $e^{+}e^{-}$ annihilation 
processes is suppressed by selecting events with low 
multiplicity as characterized by the quantity
$X_{\rm part} \equiv 
  (n_{\rm tr} + n_{\gamma})_{1} \times 
  (n_{\rm tr} + n_{\gamma})_{2}$,
where $n_{{\rm tr},j}$ and  $n_{\gamma,j}$ are the numbers 
of tracks and photons in hemisphere~$j$. We require
$X_{\rm part}\leq 25$. 
Finally, in order to eliminate
Bhabha events in which 
one or both electrons  produce a shower in 
material near the interaction region,
the acoplanarity angle $\xi$ between
the first and second highest momentum tracks is required 
to be $\xi>1^{\circ}$,
where 
%acoplanarity 
$\xi\equiv||\phi_{1} -\phi_{2}|- \pi|$ 
is defined as the two-track acollinearity in azimuth.
% this eliminates 

After applying all selection criteria,
$22.71\times 10^{6}$  $\tau^+\tau^-$-pairs survive.
The background is estimated using MC simulation.
The dominant source is from continuum processes 
$e^+e^-\rightarrow q\bar{q}\,(q=u,d,s,c)$ and amounts to~5.5\%.
Backgrounds from Bhabha 
events, $\mu^+\mu^-(\gamma)$, and two-photon 
$e^+e^-\rightarrow e^+e^- e^+e^-(\mu^+\mu^-)$ 
events are estimated to be 
0.6\%, 0.4\%, and 0.8\%, respectively.
Other sources are found to be small. These background estimates are 
checked by comparing the number of events in control samples.
The control samples for continuum, 
Bhabha\,+\,$\mu^+\mu^-$, and two-photon processes are 
high multiplicity events having $25<X_{\rm part}<30$ or
$|MM|< 0.5~\GeVCC$ or $|MM|>8.0~\GeVCC$, respectively. The differences
in event yields for these control samples and the
MC predictions (5-10\%) are included as 
systematic errors for the results discussed in latter sections.

\subsection{$\Taupipi0$ candidate selection}

Within  the $\tau^+\tau^-$-pair sample, $\Taupipi0$ decays are reconstructed 
by requiring that there be both one charged track
and one ${\pi ^{0}}$ in one hemisphere. 
The $\pi^{0}$ candidate is selected based on the normalized
invariant mass
$S_{\gamma\gamma}\equiv 
(m_{\gamma\gamma} - m^{}_{\pi^0})/\sigma_{\gamma\gamma}$,
where 
$\sigma_{\gamma\gamma}$ is the mass resolution of the $\gamma\gamma$ system.
The value of $\sigma_{\gamma\gamma}$ ranges from 0.005 GeV to 0.008 GeV,
depending on the $\pi^{0}$ momentum and polar angle.
Pairs of photons with $|S_{\gamma\gamma}|<9.0$ are considered as $\pi^{0}$
candidates.
To keep beam-related background at a negligible level, 
we require that the CM momentum of the ${\pi^0}$ be greater than 
0.25~GeV/$c$ and the photon CM energy be greater than 0.08~GeV.

\begin{figure}[t]
\begin{center}
\rotatebox{0}{\includegraphics[width=0.6\textwidth,clip]
%{./Fig/EPS2005fig/mresol.eps}}
{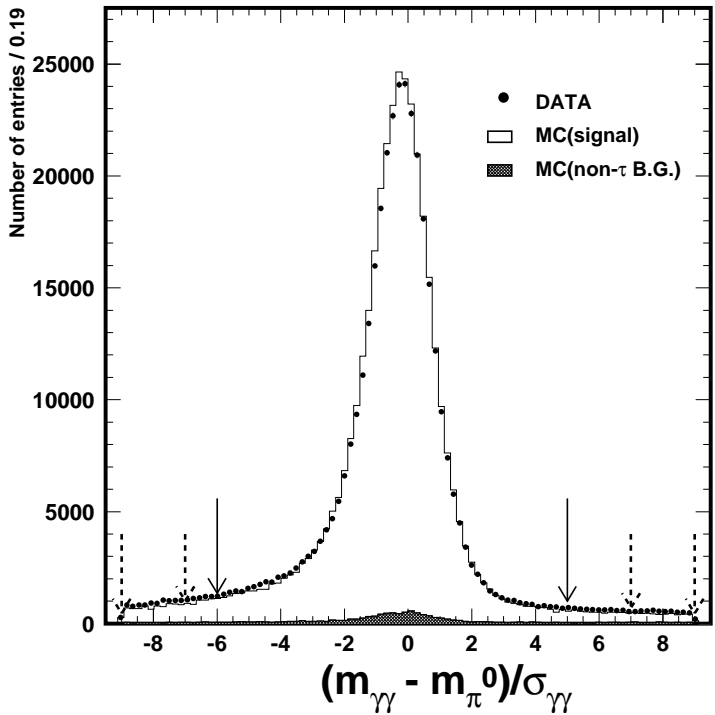}}
%
%\framebox[55mm]{\rule[-21mm]{0mm}{43mm}}
\caption{
 Normalized $\gamma\gamma$ invariant mass ($S_{\gamma \gamma}$)
spectrum in the data(points) and the $\tau^+\tau^-$ MC event(histogram),
for the sample described in the text.
 The plotted data correspond to 6.1\% of the full data used in this analysis.
The arrows indicate the signal region
$-6 < S_{\gamma \gamma} < 5$ and the sideband regions  
$9 < \left|S_{\gamma \gamma}\right| < 7$. The sideband regions 
are used to subtract fake-$\pi^{0}$ background.
}
\label{mresol}
\end{center}
\end{figure}

The distribution in the normalized di-photon invariant mass $S_{\gamma\gamma}$
for the selected $\pi^{-}\pi^{0}$ sample, where there are one charged track and one $\pi^{0}$ candidate 
in one
hemisphere, 
is shown in Fig.~\ref{mresol}.
The lower-side tail of the $S_{\gamma\gamma}$ distribution is 
primarily due to
rear and transverse leakage of electromagnetic showers out of the 
CsI(Tl) crystals and the conversion 
of the photons in the material located in front of  the crystals. 
Good agreement between data and Monte-Carlo indicates that
these effects are properly 
modeled by the Monte-Carlo simulation. 
We define the interval ${-6.0 < S_{\gamma \gamma} < 5.0}$ as the
$\pi^{0}$ signal region.
Spurious $\pi^{0}$ background 
is small and estimated from the sideband regions
${7 < \left|S_{\gamma \gamma}\right| < 9}$.
To reduce feed-down background from multi-$\pi^{0}$
decays such as 
$\tau^-\rightarrow\pi^- (n\pi^{0})\nu_{\tau}$
( $n\ge 2$),
signal candidates (in a hemisphere) are rejected 
if there are additional $\gamma$'s 
in the same hemisphere with energy greater than 0.2~GeV.

%-------------- pi0 signal ----------------------%
%------------------------------------------------------------------------%

\begin{figure}[t]
\begin{center}
\rotatebox{0}{\includegraphics[width=0.6\textwidth,clip]
%{./Fig/EPS2005fig/mono_m2pipi0_bin62_log_g5030.eps}}
%
 {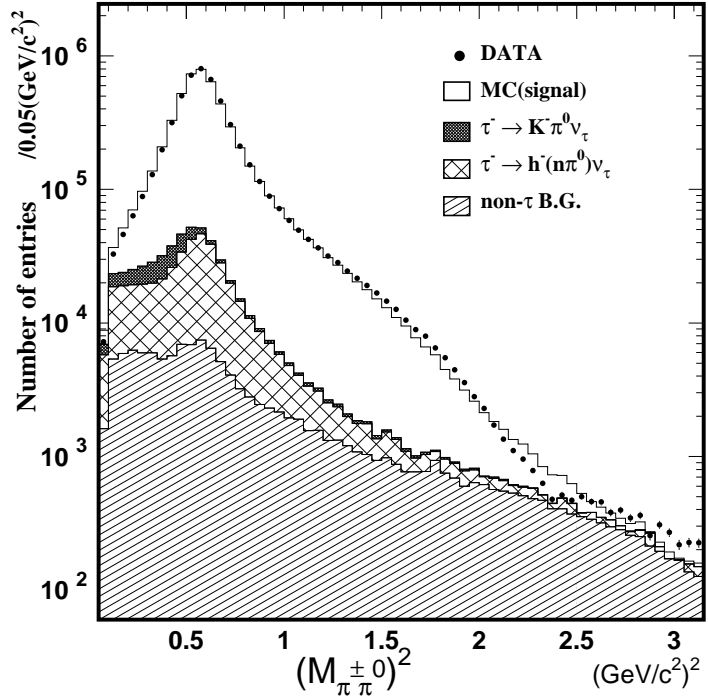}}
%
%
%\framebox[55mm]{\rule[-21mm]{0mm}{43mm}}
\caption
{
Invariant-mass-squared ($\MassSQ$) distribution for $\Taupipi0$.
The solid circles with error bars 
represent the data, and
the histogram represents MC simulation (signal\,+\,background).
The open area shows the contribution from $\Taupipi0$; 
the narrow cross-hatched 
area shows that from ${\tau^{-}\ra K^{-} \pi^{0} \nu_{\tau}}$; 
the wide cross-hatched 
area shows that from ${\tau^{-}\ra h^{-}(n\pi^{0})\nu_{\tau}}$;
and the striped area
shows that from $q\bar{q}$ continuum 
and other non-$\tau$ processes.
}
\label{pipi0_log}
\end{center}
\end{figure}

The $\pi^{-}\pi^{0}$ invariant-mass-squared ($\MassSQ$) spectrum
is obtained assuming the pion mass for the charged track; it is shown
in Fig.~\ref{pipi0_log} along with the MC prediction. To improve the
energy resolution of the $\pi^{0}$, a $\pi^{0}$ mass constraint is imposed.
%
%From this distribution one observes a spurious $\pi^{0}$ background 
%of 4--15\%,
%The amount of a spurious $\pi^{0}$ background is 4--15\%,
% depending on the $\MassSQ$ region; this background is
%subtracted using a sideband region.
The amount of a spurious $\pi^{0}$ background 
depends on the $\MassSQ$ region, varying from 4\% to 15\%.
(This is subtracted using $S_{\gamma\gamma}$ sidebands.) 
The final sample contains $5.55\times 10^{6}$ $\Tauhpi0$ candidates
after the ${\pi^{0}}$ background subtraction, where $h^{-}$ denotes
$\pi$ or $K$. This sample is 50 times larger than those of previous 
studies.
%in the previous high statistics experiments.

Feed-down background arises mainly from multi-$\pi^{0}$ modes 
such as $\tau^{-}\rightarrow \pi^{-}(n\pi^{0})\nu_{\tau}$ (5.5\%)
and $\tau\rightarrow K^{-}\pi^{0}\nu_{\tau}$ (0.48\%).
Including other modes, the total feed-down background is
$(6.0\pm 0.1)$\%.
%$5.98\pm 0.08\%$. 
The error listed includes statistical uncertainty 
as well as the uncertainty in relevant branching fractions.
Background from non-$\tau$ processes is negligible,
except that from continuum processes. 
%$e^{-}e^{+}\rightarrow q\bar{q}(q=u,d,s,c)$.
The amount of continuum background is 
estimated from MC simulation
to be $(2.45\pm 0.05)\%$. The normalization 
of the continuum MC is checked using data
in the high-mass region $\MassSQ>M^2_\tau$.

The MC simulation of $\tau$ decays is based on the TAUOLA
program~\cite{TAUOLA2004}. A small difference observed 
between data and MC 
in Fig.3 for $\MassSQ \ge 2.0 $ $\rm{(GeV/c^{2})^{2}}$ 
is attributed to the $\rho^{\prime\prime}(1700)$ resonance, which
is not included in the current TAUOLA program.
%
%-------------- pipi0 spectrum ----------------------%

\section{Measurement of the Branching Fraction}
\subsection{Formula}

The branching fraction for $\Tauhpi0$ 
($\mathcal{B}_{h\pi^0}$) is determined 
by dividing the signal yield $N_{h\pi^{0}}$ by the number of
selected generic 
$\tau^+\tau^-$-pairs $N_{\tau\tau}$:  
\begin{eqnarray}
\mathcal{B}_{h\pi^{0}}   &=& \frac{N_{h\pi^{0}}}{2 N_{\tau\tau}}\cdot
	 \frac{ (1 - b^{{\rm feed}\mbox{-}{\rm down}}-
	   b^{{\rm non}\mbox{-}\tau})}
	      { (1 - b_{\tau\tau})}  \cdot
  \left(       \frac{\epsilon_{\tau\tau}}
	     {\epsilon^{\tau}_{h\pi^{0}}}
   \right)
   \cdot \frac{1}{\epsilon^{ID}_{h\pi^{0}} }\,.
\label{br}
\end{eqnarray}
\noindent
In this formula, $b_{\tau\tau}$ is the background fraction 
in the $\tau^+\tau^-$ sample,
$\epsilon_{\tau\tau}$ is the efficiency of the $\tau^+\tau^-$-pair selection,
$\epsilon^{\tau}_{h\pi^{0}}$ is the efficiency for
$\tau^-\!\rightarrow\!h^-\pi^{0}\nu$ decays to pass 
the $\tau^+\tau^-$-pair selection,
and $\epsilon^{ID}_{h\pi^{0}}$ is the efficiency for
$\tau^-\!\rightarrow\!h^-\pi^{0}\nu$ decays
satisfying the $\tau^+\tau^-$-pair selection 
to pass the $h^-\pi^{0}$ selection.
The product $\epsilon^{\tau}_{h\pi^{0}}\cdot \epsilon^{ID}_{h\pi^{0}}$ 
is the overall detection efficiency for the $h^-\pi^{0}\nu$ final state.
The parameter $b^{{\rm feed}\mbox{-}{\rm down}}$ is the fraction of 
$h^{-}\pi^0\nu$ candidates coming from other $\tau$ decay modes, and 
$b^{{\rm non}\mbox{-}\tau}$ is the fraction coming from non-$\tau$ 
processes. In this formula, several common uncertainties such as 
that in the luminosity, that in the cross section for 
$\tau^+\tau^-$-pair production, that in the trigger efficiency, and that
in the $\tau^+\tau^-$ selection efficiency cancel in the ratio.
%of efficiencies.   
%$f_{b} = \frac{\epsilon^{\tau}_{h\pi^{0}}}{\epsilon_{\tau\tau}}$.
The value for each factor is listed in Table~\ref{tab:br} along
with the statistical error.

%------------------------ Branching fraction table -------------------------%
\begin{table}[!htb]
\renewcommand{\arraystretch}{1.4} % enlarge line spacing
\begin{center}                      
\begin{tabular}{l|c}  \hline
\hline
Parameter &\hspace{1.5cm} Values \hspace{1.5cm}   \\
\hline 
\hline    
% ${N_{\tau\tau}}$  & 
%       ${(1.378 \pm 0.001) \times 10^{6} }$ &                
%  \\
%${N_{h \pi^{0}}}$  & 
%      ${(3.29 \pm 0.005) \times 10^{5} }$  &                 
% \\
${\varepsilon_{\tau\tau}}$  &  
 ${30.81 \pm 0.05~\%}$  \\ 
${\varepsilon^{\tau}_{h\pi^{0}}}$ &  ${34.26\pm 0.07 ~\%}$  \\ 
${\displaystyle f_{b} = 
	     \frac { \varepsilon^{\tau}_{h\pi^{0}} }
			 { \varepsilon_{\tau\tau} }   }$ &
	      ${1.112 \pm 0.003 }$  \\ 
${\varepsilon_{h\pi^{0}}^{ID}}$ &
	      ${42.62 \pm 0.13 ~\%}$ \\ 
${b_{\tau \tau}}$ &
	      ${7.66 \pm 0.03~\%}$  \\ 
${ b^{{\rm feed}\mbox{-}{\rm down}}_{h\pi^{0}} }$ & 
	      ${5.98 \pm 0.08 ~\% }$  \\
${ b^{{\rm non}\mbox{-}\tau}_{h\pi^{0}} }$  &
	      ${2.45 \pm 0.06 ~\% }$  \\
\hline
\hline
\end{tabular}
\caption{ Values of parameters used for the 
branching fraction measurement along with statistical errors.}
\label{tab:br}
\end{center}
\end{table}

\subsection{Systematic uncertainty}

The sources of systematic uncertainty are listed in Table~\ref{tab:br_sys}. 
The uncertainty in the tracking efficiency is estimated using  
$D^{*\,+}\rightarrow D^0\pi^+\rightarrow K^-\pi^+\pi^+$
% and 
%$\eta\rightarrow\pi^+\pi^-\pi^0$ 
decays to be 
1\% per track.
A large part of this uncertainty cancels in the ratio  of
Eq.(\ref{br}); the resulting uncertainty  from this source is
$\Delta \mathcal{B}_{h\pi^{0}}=0.12$ \%.
%1.0--1.5\%.
The $\gamma/\pi^{0}$ detection efficiency is  
%calibrated 
obtained from the ratio of
$D^0\rightarrow K^-\pi^+\pi^{0}$
to $D^0\rightarrow K^-\pi^+$ decays, in 
which the branching fractions are known
precisely. The uncertainty is estimated to be
$\pm 1.7\%$ for a $\pi^{0}$ momentum less than 1.0~GeV/$c$.
As a consistency check, the branching fraction is re-measured 
after changing the photon threshold from 
0.05~GeV to 0.10~GeV; the difference 
in $\mathcal{B}^{}_{h\pi^0}$ is only 
0.10\%.
The uncertainty in background in 
the non-$\tau$ sample
$\delta b_{h\pi^{0}}^{\rm non-\tau}$
is estimated from the control sample
above the $\tau$ mass region,
 while the uncertainty
in  feed-down
%$\Tauhpi0$ 
background $\delta b_{h\pi^{0}}^{\rm feed-down}$
 is obtained 
from the uncertainty in 
$\tau^{-}\!\rightarrow\!h^{-}(n\pi^{0})\nu_{\tau}$ and
$\tau^-\!\rightarrow\!K^{-}\pi^{0}\nu_{\tau}$ 
branching fractions. 
%The largest source of systematic uncertainty arises from
%low momentum photons.

\begin{table}[htbp!]
\begin{center}
\begin{tabular}{l|c} 
\hline 
\hline
Source of  uncertainty & $\Delta \mathcal{B}_{h\pi^{0}}$ (\%)  \\
\hline
\hline
%    $\gamma$ detection &$\pm1.25$ & \\
Tracking efficiency & 0.12  \\
$\pi^0/\gamma$ efficiency & 0.25  \\
Background for $\tau^+\tau^-$  &   0.09 \\
Feed-down background for $\Tauhpi0$   & 0.04  \\
Non-$\tau$ background for $\Tauhpi0$  & 0.05   \\
$\gamma$ veto  &  0.05    \\
Trigger            & 0.08 \\
%(8) Exp. dependence    &  0.08  \\
MC statistics &   0.04  \\
\hline
\hline
%\hspace{1cm} 
Total &  0.31 \\
\hline
\hline
\end{tabular}
%\tiny
\end{center}
\caption{ Systematic uncertainties for the $\Tauhpi0$ branching fraction.
}
\label{tab:br_sys}
\end{table}

The veto of additional $\gamma$'s is required in the event
selection to reduce background from multi-$\pi^0$ decay channels. 
However, it also vetoes signal if there are photons radiated in 
the initial or final state. In addition, photon candidates
 can 
also appear due to electromagnetic shower fragments and/or 
mis-reconstructed of electrons.
The uncertainty from these sources is estimated by 
changing the veto threshold by 
%$\bigtriangleup E_{\gamma}^{th} = 
$\pm 0.1$~GeV; the resulting change in 
$\mathcal{B}^{}_{h\pi^0}$ 
is only
%is $\Delta \mathcal{B} = 
$\pm 0.05$\%.
%%%%%%%%%%%%%%%%%%%%%%%%%%%%%%%%%%%%%%%%%%%%%%%%%%%%%%%%%%%%%%%%%%%%%%
%The trigger efficiency has been monitored by the trigger group
%continuously and
%its results are summarized in the series of the Belle notes.
%The largest uncertainty of the trigger comes from the uncertainty of the
%z-trigger efficiency of the CDC and the TOF efficiencies.
%There are triggered both track trigger and energy triggers.
%%%%%%%%%%%%%%%%%%%%%%%%%%%%%%%%%%%%%%%%%%%%%%%%%%%
Signal events are flagged by several trigger conditions
that require two or more CDC tracks with associated TOF hits, 
ECL clusters, or a significant sum of energy in the ECL. 
This redundancy 
%of the trigger condition 
allows one to monitor the efficiency of each trigger requirement.
The uncertainty arising from the trigger is estimated by
assuming there is a $\pm 3$\% uncertainty in the track and 
energy trigger efficiencies, which is the maximum variation 
%of the track and energy trigger efficiencies 
measured during experimental running. The resulting uncertainty 
on $\mathcal{B}_{h\pi^{0}}$ is small (0.08\%) since the 
$\tau^+\tau^-$ trigger efficiency is high (97\%).

\subsection{Results}

Inserting all values into Eq.~(\ref{br}) gives
%Taking into account all factors, the branching fraction for $\Tauhpi0$ 
%is measured to be
\begin{eqnarray}
\mathcal{B}_{h\pi^{0}} & = & (25.60\,\pm \,0.04\,\pm\,0.31)\%\,,  
\end{eqnarray}
\normalsize
where the first error is statistical and the second is systematic.
Subtracting the small kaon-channel branching fraction listed in the 
PDG~\cite{PDG2004} [$\mathcal{B}_{ K^-\pi^0}=(0.45\pm 0.03)\%$] 
gives a $\Taupipi0$ branching fraction of
\begin{eqnarray}
\mathcal{B}_{\pi\pi^{0}} & = & (25.15\,\pm\,0.04\,\pm\,0.31)\%\,.
\end{eqnarray}
This result is in good agreement with previous measurements,
as shown in Table~\ref{tab:br_comp}. Our statistical error is 
significantly lower than those of the other measurements,
while our systematic error is similar to those of the others 
(except for the ALEPH result).
%Our systematic error is similar 
%to those of previous measurements, with the exception of
%that from the ALEPH experiment.

\begin{table}[htbp!]
\begin{center}
\begin{tabular}{l|c |c} 
\hline 
\hline 
Experiment & $\mathcal{B}_{h\pi^{0}}(\%)$ & Reference \\ 
\hline
\hline
OPAL & $25.89 \pm 0.17 \pm 0.29$  & \cite{OPAL98M}   \\
%\hline
%ALEPH & $25.76 \pm 0.15 \pm 0.13$ \\
ALEPH & $25.924 \pm 0.097 \pm 0.085$ & \cite{ALEPH05}   \\
%\hline
L3 & $25.05 \pm 0.35 \pm 0.50$   &\cite{L395}       \\
%\hline
CLEO & $25.87 \pm 0.12 \pm 0.42$ & \cite{CLEO94} \\
\hline
This work & $25.60 \pm 0.04 \pm 0.31$       \\
\hline
\hline
\end{tabular}
\end{center}
\caption{ 
Branching fractions for $\Tauhpi0$ measured 
by various experiments. 
%The OPAL, L3, and CLEO results
%are taken from the PDG~\cite{PDG2004}. 
%The ALEPH result is taken from
%Ref.~\cite{ALEPH05}.
}
\label{tab:br_comp}
\end{table}

%%The accuracy is limitted by the uncertainty of the $\gamma/\pi^{0}$
%detection efficiency.

\section{Measurement of the Mass Spectrum}

In order to obtain the true $\pi^{-}\pi^{0}$ mass spectrum,
one must correct for (1) background, (2) smearing due to finite 
resolution and radiative effects, and (3) mass-dependent acceptance.

\subsection{Background Correction}

As noted earlier, background entering the $\Taupipi0$ sample 
is small. The sidebands of the $M_{\gamma\gamma}$ distribution 
are used to estimate the fake $\pi^{0}$ contribution.
% this is found to be
%
%8--10\% on average depending on the experimental conditions.
This background dominates at values of $\MassSQ$ less 
than about $0.25~\GeVcc2$.

\begin{figure}[t]
\begin{center}
\rotatebox{0}{\includegraphics[width=0.6\textwidth,clip]
%{./Fig/EPS2005fig/g5030/m2pp_without-bg_62_g5030.eps}}
{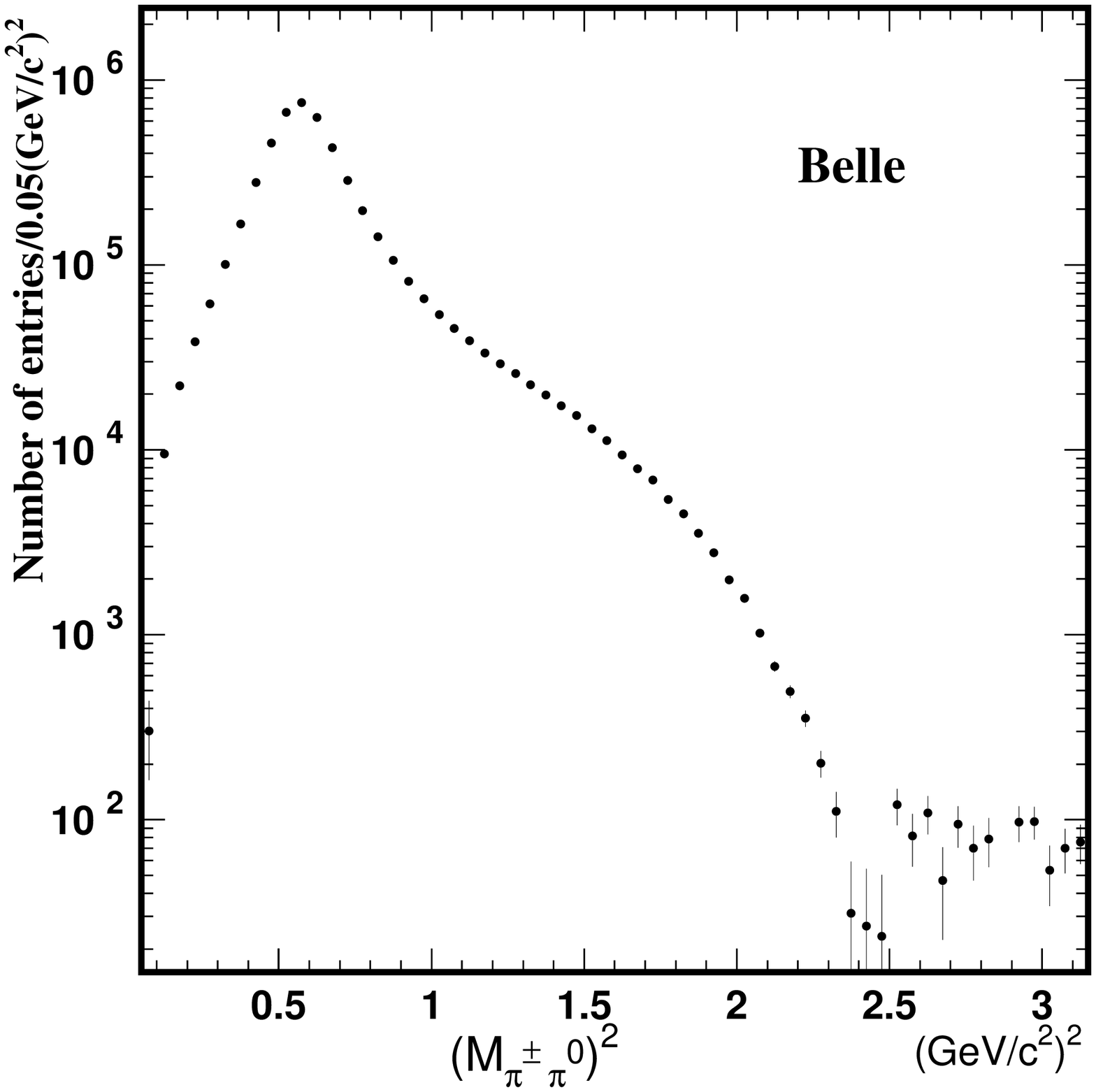}}
%
%\framebox[55mm]{\rule[-21mm]{0mm}{43mm}}
\caption
{
Invariant-mass-squared ($\MassSQ$) distribution for $\Taupipi0$
after background subtraction.
}
\label{pipi0_aftsub}
\end{center}
\end{figure}

As seen in Fig.~\ref{pipi0_log}, feed-down background arises 
from $\tau^{-}\rightarrow h^{-}(n\pi^{0})\nu_{\tau}$ 
and $\tau^{-}\rightarrow K^{-}\pi^{0}\nu_{\tau}$ decays;
both backgrounds dominate at low values of $\MassSQ$.
In the high mass region, continuum background dominates.
For this analysis we did not use information 
from particle identification~(PID) detectors 
to separate charged pions from kaons, as the
feed-down background is dominated by 
$\tau^{-}\rightarrow h^{-}(n\pi^{0})\nu_{\tau}$ 
rather than $\tau^{-}\rightarrow K^{-}\pi^{0}\nu_{\tau}$.
%The reasons are two-hold. One is to avoid  the additional uncertainty
%coming  from PID requirements. The second
% is that
%the feed-down background 
The $\MassSQ$ distribution after subtracting this
background is shown in Fig.~\ref{pipi0_aftsub}. 
%The spectrum ranges five orders of  magnitude.
%with small errors in the wide mass region.

\subsection{Acceptance  Corrections}

The detector effects include $\MassSQ$-dependent 
acceptance and bin-by-bin migration caused by the finite mass resolution. 
We correct for these effects by performing an unfolding procedure. The 
unfolding program used is that employed by the ALEPH experiment~\cite{SVD}.
%developed by  A.~H\"{o}cker 
In this program, the unfolding is based on the
Singular-Value-Decomposition (SVD) method~\cite{SVD}, 
in which the acceptance matrix is inverted 
%the well-known problem of inverting the acceptance matrix
%is solved 
by limiting the number of singular values to
only those that are statistically significant.
The output of the program is the unfolded distribution and 
its covariance matrix.
%%%%%%%%%%%%%%%%%%%%%%%%%% 
%singular-value-decomposition(SVD) method. 
%One important feature of the SVD unfolding method is that
% an optimization(regularization) procedure is  clearly defined.
%  
%${A.H\ddot{o}ker}$, ${V.Kartvelishvili}$, ${N.I.M.}$ \\
%%%%%%%%%%%%%%%%%%

\begin{figure}[t]
\begin{center}
\rotatebox{0}{\includegraphics[width=0.6\textwidth,clip]
%{./Fig/EPS2005fig/accept_con_g5030.eps}}
 {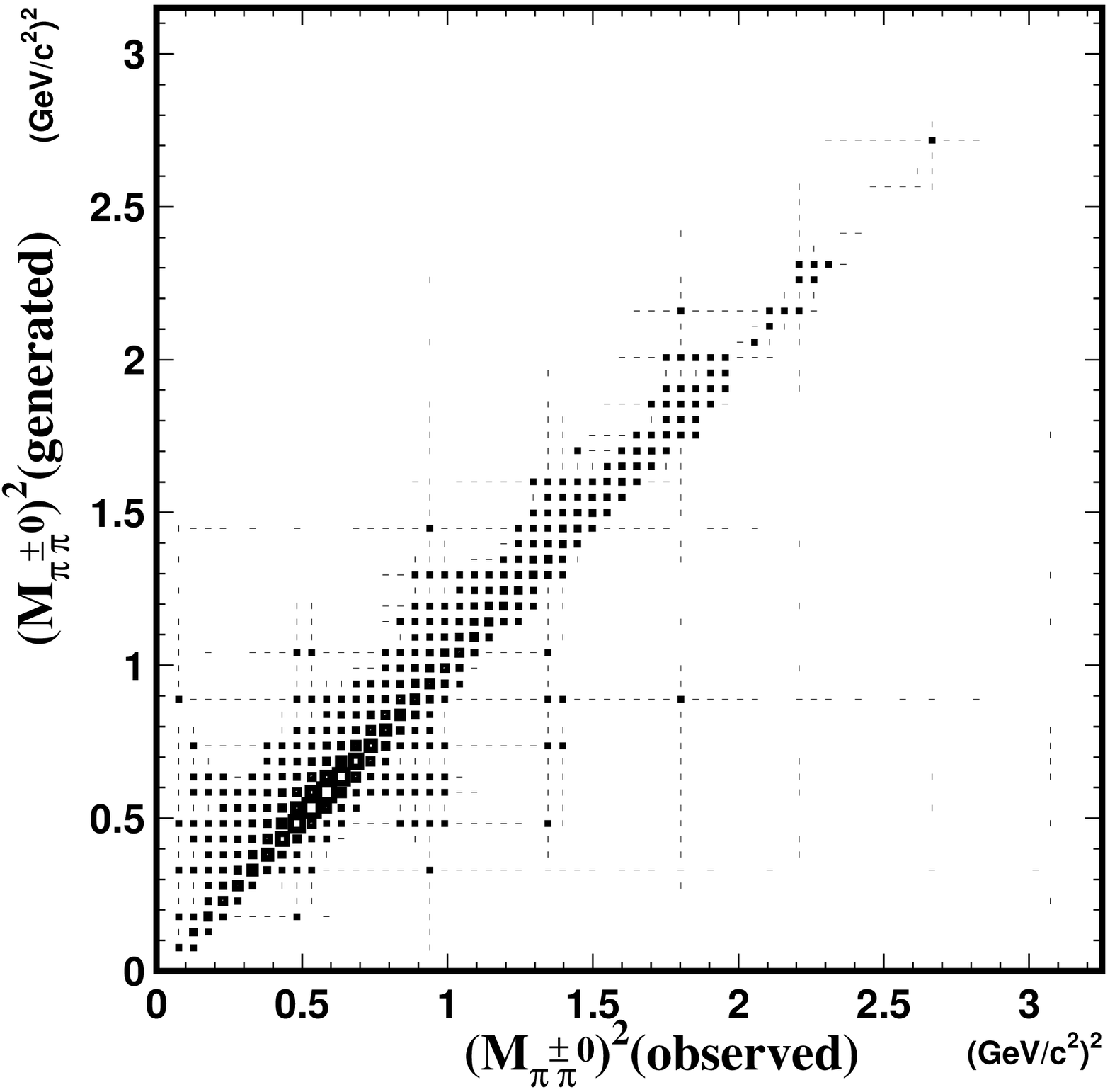}}
\caption
{
Correlation between $\MassSQgen$ (vertical axis) and 
$\MassSQobs$ (horizontal axis), the generated and
observed invariant masses squared of the $\pi^-\pi^{0}$ 
system in $\Taupipi0$ decay.
%Note that the logarithmic scale is used in
%the contour lines.
}
\label{cor_matrix}
\end{center}
\end{figure}

The correlation between the generated quantity 
$\MassSQgen$ and the measured one $\MassSQobs$ 
is shown in Fig.~\ref{cor_matrix}. The figure 
shows a clear correlation between 
$\MassSQgen$ and $\MassSQobs$.
The resolution in $\MassSQ$ is 0.005~$\GeVcc2$
in the low-mass region and 
0.030~$\GeVcc2$ in the 
high-mass region; thus the bin size chosen is
$\Delta M^{2} = 0.050~\GeVcc2$ so that the off-diagonal 
components of the acceptance matrix are small.

%-------------- correlation matrix ----------------------%
%------------------------------------------------------%

\begin{figure}[t]
\begin{center}
\rotatebox{0}{\includegraphics[width=0.6\textwidth,clip]
%
%{./Fig/EPS2005fig/acceptance2_g5024.eps}}
%
  {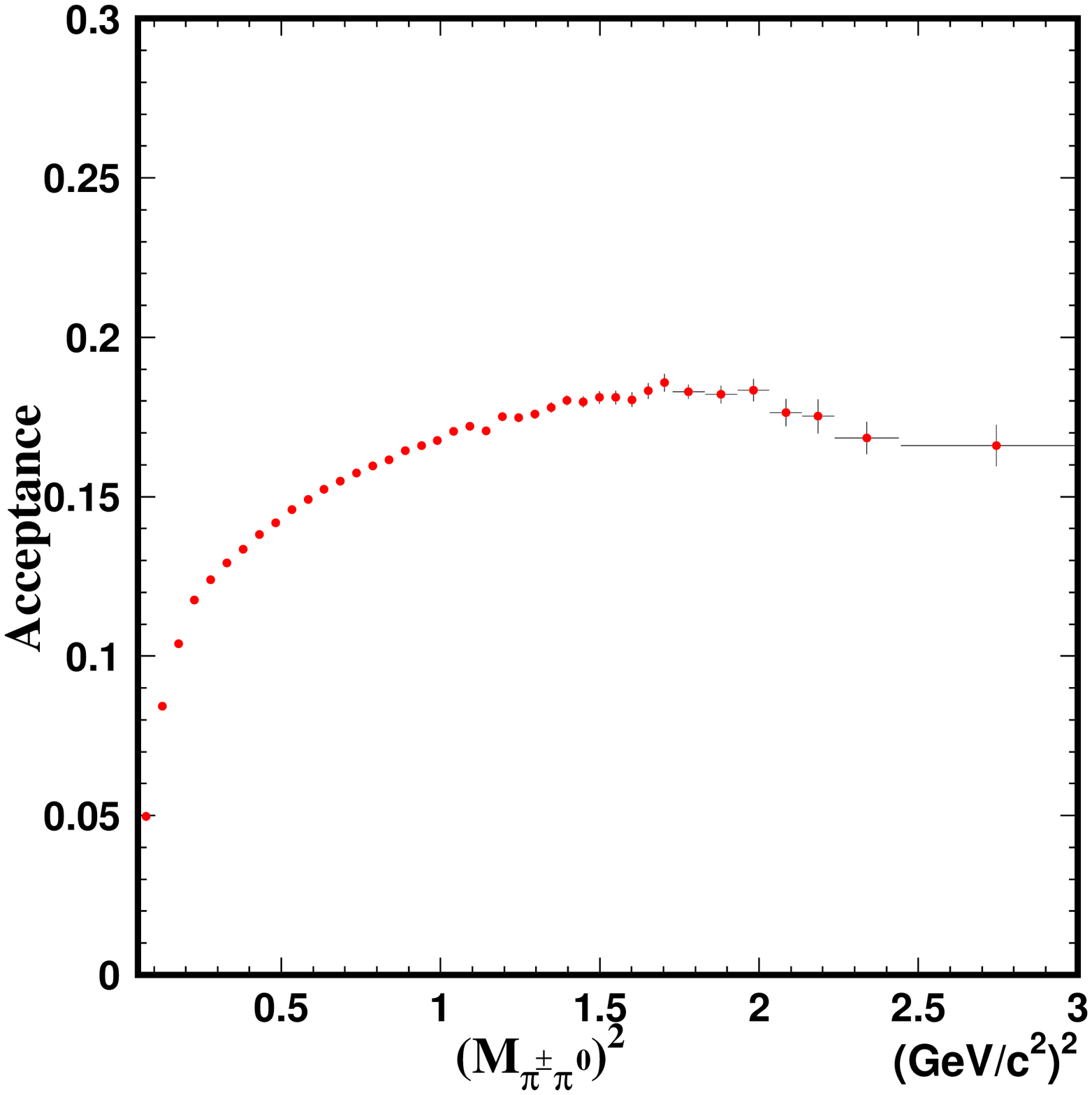}}
%
%\framebox[55mm]{\rule[-21mm]{0mm}{43mm}}
\caption
{Acceptance as a function of $\MassSQgen$, as determined
from $\Taupipi0$ MC simulation.  }
\label{acceptance}
\end{center}
\end{figure}

The acceptance as a function of $\MassSQgen$ is shown in 
Fig.~\ref{acceptance}. The acceptance varies smoothly 
and its average value is~17\%. It decreases at low values
of $\MassSQgen$ due to the overlap of $\gamma$ clusters
with the $\pi^{-}$ track projection at the calorimeter.

%-------------- acceptance ----------------------%
%-------------------------------------------------%

\subsection{Results}

\begin{figure}[t]
\begin{center}
\rotatebox{0}{\includegraphics[width=0.6\textwidth,clip]
%
%{./Fig/EPS2005fig/g5030/stat+sys2/bwgs3p_fit_emat_g5030_set40.eps}}
%
{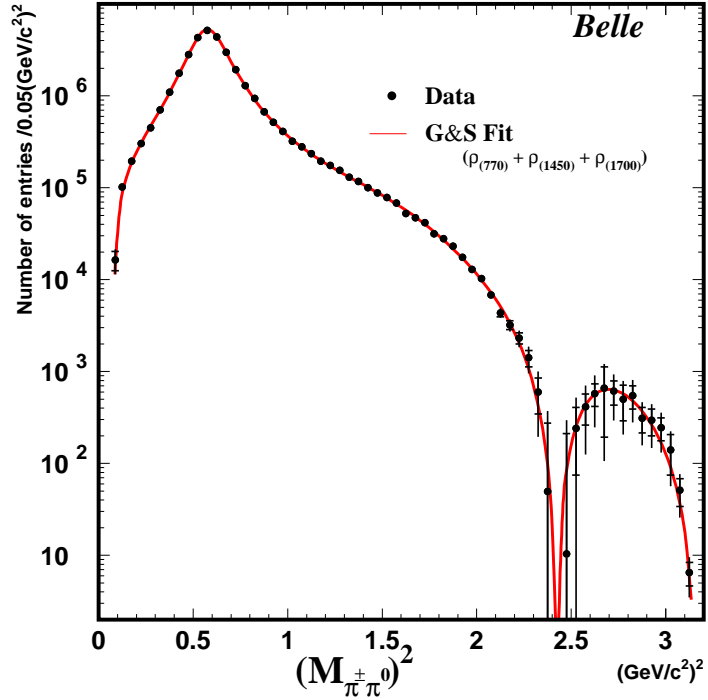}}
\end{center}
\caption
{ Fully-corrected $M^2_{\pi\pi^0}$ distribution
for $\Taupipi0$. 
The solid curve is the 
result of a fit to the Gounaris-Sakurai model with 
$\rho(770)$, $\rho'(1450)$, and $\rho''(1700)$ resonances.
All parameters are floated.
}
\label{unfold_pipi0}
\end{figure}

The unfolded $s= M^2_{(\pi\pi^0\,{\rm unf.})}$ spectrum
$dN/ds$ is shown in Fig.~\ref{unfold_pipi0}. 
%Here $s$ is the invariant-mass squared of the $\pi^{-}\pi^{0}$:
% $s=\MassSQ$.
The square roots of the diagonal components of the covariance
matrix are used as the errors.
The $\rho$ peak and the shoulder due to the $\rho'(1450)$
are clearly visible. The dip at $s\approx 2.5~\GeVcc2$ is
caused by destructive interference between the $\rho'(1450)$ 
and $\rho''(1700)$ resonances.

%------------------- pipi0 mass spectrum (unfold) --------------------%
%----------------------------------------------------------------------%

%\subsection{Extraction of  resonance parameters }

To obtain parameters for the $\rho$, $\rho'$ and $\rho^{''}$ 
resonances, a $\chi^{2}$ fit using Breit-Wigner functions 
is performed. Since the unfolded mass spectrum has bin-by-bin 
correlations, the off-diagonal components of the covariance matrix
$X$ are included in the $\chi^{2}$ evaluation:
\begin{equation}
\chi^{2}= \sum_{i,j} \left( y _{i} - f(s_{i}; \alpha) \right)
	(X^{-1})_{ij}
	\left( y _{j} - f(s_{j}; \alpha) \right)\,,
\label{eq:chi2def} 
\end{equation}
where $y_{i}$ is the measured value at the $i$-th bin,
$f(s; \alpha)$ is the value of the function for parameters $\alpha$,
and $(X^{-1})_{ij}$ is the inverse of the covariance matrix. 

In the fit, the $s$ dependence of the decay rate is given by 
Eq.~(\ref{eq:tauspec}). The pion form factor in Eq.~(\ref{eq:pionform})   
is parametrized with Breit-Wigner functions corresponding to the 
$\rho$, $\rho^{\prime}(1450)$, and $\rho^{\prime\prime}$(1700) 
resonances:
%%%%%%%%%%%%%%%%%
%
%-- BW formula --%
%\begin{equation}
% \frac{dN}{dM^{2}} = A
%                     \left( 1 - \frac{M^{2}}{M_{\tau}^{2}} \right) ^{2}
%                     \left( 1 + \frac{2M^{2}}{M_{\tau}^{2}} \right) \cdot
%                     |F_{\pi}(M^{2})|^{2}
%                     \beta_{\pi}^{3}
%\end{equation}
%%%%%%%%%%%%%%%%%%
\begin{equation}
F_{\pi}(s) = \frac{1}{1 + \beta +\gamma } 
	  (BW_{\rho} + \beta \cdot  BW_{\rho^{\prime}} 
 +\gamma \cdot BW_{\rho^{\prime\prime}})\,,
\end{equation}
where the parameters $\beta$ and $\gamma$
(denoting the relative size of the two resonances) are 
in general complex.
% $\beta=|\beta|e^{i\phi}$.
We use the Gounaris-Sakurai\,(GS) model~\cite{GS} for the
Breit-Wigner shape:
\begin{equation}
BW_{i}^{GS} = \frac {M_{i}^2  + d \cdot M_{i}\Gamma_{i}(s) }
	   {(M_{i}^{2} - s) + f(s) - i \sqrt{s} \Gamma_{i}(s)}\,,
\end{equation}
\noindent
with an energy-dependent width
\begin{equation}
\Gamma_{i}(s) = \Gamma_{i}  \left( \frac{M_{i}^{2}}{s}\right)
\left( \frac{k(s)}{k(M_{i}^2)}\right)^{3}\,.
\end{equation}
Here, $ k(s)  =  \frac{1}{2} \sqrt{s} \beta_{-}(s)$ is the pion
momentum in the $\pi^{-}\pi^{0}$ rest frame.
The functions $f(s)$ and $h(s)$ are defined as

\begin{eqnarray}
f(s) & = & \Gamma^{}_i\,\frac{M^2_i}{k^3(M^2_i)} 
  \left[\, 
     k^2(s) \left( h(s) -h(M_{i}^2) \right)
     + (M_{i}^2 - s) k^2(M_{i}^2) 
	\left.\frac{dh}{ds}\right|_{s=M^2_{i}} 
  \,\right]  
\label{eq:dif} \\
& & \nonumber \\  
h(s) & = & \frac{2}{\pi} \frac{k(s)}{\sqrt{s}} 
ln \frac{\sqrt{s} + 2k(s)}{2m_{\pi}}\,,
\end{eqnarray}
\noindent
with  
$ \left. dh/ds\right|_{M_{i}^{2}} = 
h(M_{i}^{2})  \left [  
		 \left( 8k^2(M_{i}^{2}) \right)^{-1}
		  - (2 M_{i}^{2})^{-1}    \right ]
+ (2 \pi M_{i}^{2})^{-1}
$
and  
%The parameter $d$ is defined such that
%$BW^{GS}_{i}=1$ at $s=0$:

\begin{equation}
    d = \frac{3}{\pi} \frac{m_{\pi}^2} {k^2(M_{i}^2)} 
   ln \frac{M_{i} + 2 k(M_{i}^2)} {2 m_{\pi}}
		 +
  \frac{M^{}_i} {2 \pi k(M^2_{i})} -
  \frac{m_{\pi}^2 M^{}_i} {\pi k^3(M^2_{i})}\,.          
\end{equation}
\noindent
There are ten parameters in this formula: 
the masses~($M_{i}$) and the widths~($\Gamma_{i}$) for the 
$\rho,\,\rho^{\prime}$,\, and $\rho^{\prime\prime}$ resonances,
their relative amplitudes $|\beta|$,\,$|\gamma|$,\, and their phases
$\phi_{\beta}$ and $\phi_{\gamma}$.

%\begin{table}
%\begin{tabular}{c|c}  
%\hline 
%\hline 
%Parameter &  Fit result \\
%\hline 
%\hline 
%  $M_{\rho}$ & $774.6\pm 0.2\pm 0.3  ~~{\rm MeV/c^{2}}$ \\
%  $\Gamma_{\rho}$ & $150.7\pm 0.3\pm 0.3  ~~{\rm MeV}$ \\
%  $M_{\rho^{\prime}}$ & $1331\pm 10\pm 2.8  ~~{\rm MeV/c^{2}}$ \\
%  $\Gamma_{\rho^{\prime}}$ & $468\pm 32\pm 6.5  ~~{\rm MeV}$ \\
%  $\beta$ & $0.086 \pm 0.008 \pm 0.002$ \\
% $\phi_{\beta}$ & $121.7 \pm 5.4\pm 0.5$ \\
% $M_{\rho^{\prime\prime}}$ & $1591 \pm 9.3 ~~{\rm MeV/c^{2}}$ \\
%  $\Gamma_{\rho^{\prime\prime}}$ & $257\pm 16\pm 58  ~~{\rm MeV}$ \\
%  $\gamma$ & $0.059 \pm 0.009 \pm 0.007$ \\
% $\phi_{\gamma}$ & $-67.3 \pm 11\pm 5.2$ \\
%\hline
% $\chi^{2}/{\rm d.o.f}$ & 73/51 \\
%
%   $M_{\rho}$ &  $774.6\pm 0.2\pm 0.3  ~~{\rm MeV/c^{2}}$ \\
%  $\Gamma_{\rho}$ &  $150.5\pm 0.2\pm 0.3   ~~{\rm MeV}$ \\
%  $M_{\rho^{\prime}}$ & $1340\pm 9\pm 2.8  ~~{\rm MeV/c^{2}}$ \\
%  $\Gamma_{\rho^{\prime}}$ & $497\pm 24\pm 6.5   ~~{\rm MeV}$ \\
%  $|\beta|$ & $0.096 \pm 0.006 \pm $  0.002\\
% $\phi_{\beta}$ & $121.01 \pm 4.0\pm  0.5$  ~~deg.\\
% $M_{\rho^{\prime\prime}}$ & $1604 \pm 6\pm 9  ~~{\rm MeV/c^{2}}$ \\
%  $\Gamma_{\rho^{\prime\prime}}$ & $268\pm 17\pm 58  ~~{\rm MeV}$ \\
%  $|\gamma|$ & $0.068 \pm 0.011 \pm 0.007 $ \\
% $\phi_{\gamma}$ & $-62.9 \pm 3.9\pm 5.2 $ ~~deg. \\
%\hline
% $\chi^{2}/{\rm d.o.f}$ & 127/51 \\\hline
%\hline
%  \end{tabular}
%  \caption{
%      Results of fitting the $M^2_{\pi\pi^0}$ distribution 
%      for $\Taupipi0$ to the Gounaris-Sakurai model with 
%      $\rho(770)$, $\rho'(1450)$, and $\rho''(1700)$ resonances. }
%  \label{tab:fit_param}
%\end{table}

\begin{table}
\begin{tabular}{l|c|c}  
\hline 
\hline 
Parameter &  Fit result & Fit result\\
  &  (all free)  & (fixed $\phi_{\gamma}$)   \\
\hline 
\hline 
$M_{\rho}$~~$({\rm MeV/c^{2}})$    &  $774.6\pm 0.2\pm 0.3$  
&  $774.3\pm 0.2\pm 0.3$   \\
$\Gamma_{\rho}$~~$({\rm MeV})$ &  $150.6\pm 0.3\pm 0.5$   
& $150.0\pm 0.3\pm 0.5$     \\
$M_{\rho^{\prime}}$ ~~$({\rm MeV/c^{2}})$  & $1336\pm 12\pm 23$  
&   $1436\pm 15\pm 23$    \\
$\Gamma_{\rho^{\prime}}$ ~~$({\rm MeV})$  & $471\pm 29\pm 21$   
   & $553\pm 31\pm 21$     \\
$|\beta|$ & $0.090 \pm 0.009 \pm   0.013$
   &  $0.161 \pm 0.020 \pm   0.013$ \\
$\phi_{\beta}$ ~~(degree)  & $123.7 \pm 5.0\pm  7.0$  
  & $149.1 \pm 2.4\pm  7.0$  \\
$M_{\rho^{\prime\prime}}$ ~~$({\rm MeV/c^{2}})$  & $1600 \pm 13\pm 4$  
  & $1804 \pm 16\pm 4$  \\
$\Gamma_{\rho^{\prime\prime}}$ ~~$({\rm MeV})$  & $255\pm 19\pm 79$  
&  $567\pm 81\pm 79$  \\
$|\gamma|$ & $0.062 \pm 0.015 \pm 0.015$  
& $0.136 \pm 0.024 \pm 0.015$   \\
$\phi_{\gamma}$ ~~(degree) & $-64.1\pm 7.9$ 
& $[0] $  \\
\hline
$\chi^{2}/{\rm d.o.f}$ & 55/51 &  94/52   
\\
\hline
\hline
\end{tabular}
\caption{
Results of fitting the $M^2_{\pi\pi^0}$ distribution 
for $\Taupipi0$ to the Gounaris-Sakurai model with 
the $\rho(770)$, $\rho'(1450)$, and $\rho''(1700)$ resonances.
The results for two cases, all parameters floated (the second column) 
and fixed $\phi_{\gamma}$ (the third column) are shown.
For both cases,
the first error is statistical and the second one is systematic. 
The systematic errors include the uncertainty of the backgrounds, 
unfolding, as well as the uncertainty of the photon energy scale.
}
\label{tab:fit_param}
\end{table}

All parameters are floated in the fit. When evaluating the $\chi^{2}$, 
the 1\% systematic uncertainty resulting from the unfolding procedure 
is included in the diagonal part of the covariance matrix.
This uncertainty is estimated by applying the same unfolding procedure 
to MC events and comparing the unfolded spectrum with the original.  
The result of the fit is shown in Fig.~\ref{unfold_pipi0}
as the solid line; the values obtained for the parameters 
are listed in Table~\ref{tab:fit_param}.
The results are compared with the previous ALEPH measurements in
Table~\ref{tab:fit_result}.
\begin{table*}
\begin{tabular}{l|c|c|c}  
\hline \hline
Parameter & 
This work  & This work & ALEPH($\tau$)    \\
 &
(fixed $\phi_{\gamma}$) &
(fixed $M_{\rho^{\prime\prime}},
 \Gamma_{\rho^{\prime\prime}},\phi_{\gamma}$)    \\
\hline
\hline 
${M_{\rho^{-}} ~~({\rm MeV}/c^{2})}$   &
${774.3 ~ \pm 0.2}$  & $773.9~\pm 0.1$ 
& ${775.5 ~ \pm 0.7}$    \\
${\Gamma_{\rho^{-}} ~~({\rm MeV})}$   & 
${150.0 ~ \pm 0.3}$  &   $ 150.8~\pm 0.3 $
&${149.0 ~\pm 1.2}$  \\
${M_{\rho^{\prime}} ~~({\rm MeV}/c^{2})}$   & 
${1436 ~ \pm 15}$   &  $ 1395 ~\pm 4 $
&${ 1328 ~ \pm 15 }$   \\
${\Gamma_{\rho^{\prime}} ~~({\rm MeV})}$   & 
${553 ~ \pm 31}$   &  $411 ~\pm 9$
&${468 ~\pm 41}$   \\
${|\beta|}$  &
${0.161 ~ \pm 0.020}$ &   $ 0.095~\pm 0.02$
&${0.120 ~\pm 0.008}$     \\
${\phi_{\beta}}$ ~~(degree)  &
${149.1 ~ \pm 2.4}$   &$161~\pm 2.0$
&${153 ~\pm 7}$             \\
${M_{\rho^{\prime\prime}} ~~({\rm MeV}/c^{2})}$   & 
$1804 \pm 16$   & [1713]
&  [1713]    \\
${\Gamma_{\rho^{\prime\prime}} ~~({\rm MeV})}$   & 
${567 ~ \pm 81}$  & [235]  
& [235]       \\ 
${|\gamma|}$  &
${0.136 ~ \pm 0.024}$  & $ 0.045 \pm 0.002$  
&${0.023 ~\pm 0.008}$     \\
${\phi_{\gamma}}$ ~~(degree)  &
 [0]   &   [0]
& [0]            \\
%   \hline
${\chi ^{2} /{\rm (d.o.f)}}$   &
94 / 52      &  134/54
   &    119 / 110       \\
Reference         &  &
& ~\cite{ALEPH05}   \\
\hline 
\hline
\end{tabular}
\caption{ Comparison of our fit results for 
the $\rho(770)$, $\rho'(1450)$, and $\rho''(1700)$ parameters with those
obtained by the ALEPH  experiment\protect{\cite{ALEPH05}}. 
The numbers in brackets indicate
the values fixed in the fit.} 
\label{tab:fit_result}
\end{table*}
%--------------------------------------------------------------------------% 
%
%The results are compared with previous ALEPH measurements in 
%
%
In the table, the first error is statistical and the second one is
systematic.

The main sources of systematic uncertainty are 
the photon energy scale, the unfolding procedure, 
and the background subtraction.
The uncertainty in the $\rho$ mass (0.3~MeV) is 
mainly due to the uncertainty in the photon energy scale.  
The uncertainty in background dominates for the $\rho''$ parameters. 
Our result for the mass of the
$\rho$ resonance agrees well 
with the 
ALEPH~\cite{ALEPH05} and CLEO\cite{CLEO2000} results.

As can be seen from the second and third columns 
of Table~\ref{tab:fit_result},
 where 
the interference angle $\phi_{\gamma}$ or three parameters
 $M_{\rho^{\prime\prime}},\Gamma_{\rho^{\prime\prime}},\phi_{\gamma}$
are fixed, respectively,
  as in the previous  ALEPH~\cite{ALEPH05} fit,
the values for $\rho^{\prime}$ and $\rho^{\prime\prime}$  resonance parameters
are quite sensitive to the values of other parameters fixed in the fit.
%have a strong correlation.
%The values of $\chi^{2}$  indicate that the fit is poor for these fixed 
%when the
%parameters are fixed to the values used in the previous fit.
% with the parameters of the $\rho^{\prime\prime}$
%resonances. 
%Ultimately, it is expected that
% one needs data in the high mass region above $m_{\tau}$ 
%to determine the $\rho^{\prime}$ and $\rho^{\prime\prime}$ resonance
% parameters precisely.

%%%%%%%%%%%%%%%%%%%%%%%%%%%%%%%%%%%%
%In the meeting, the importance of the  possible mass and width difference
%between  $\rho^{-}$ in $\tau$ data and  $\rho^{0}$ in $e^+e^-$ data
%are emphasized~\cite{JEG2003,DASIGHAD}.
%In the table, only statistical errors are shown.
%%%%%%%%%%%%%%%%%%%%%%%%%%%%%%%%%%%%%

\begin{figure}[t]
\begin{center}
\rotatebox{0}{\includegraphics*[width=0.6\textwidth,clip]
%
%{./Fig/EPS2005fig/g5030/stat+sys2/correct_pionf3_fit_emat_g5030_set40_c.eps}}
%
{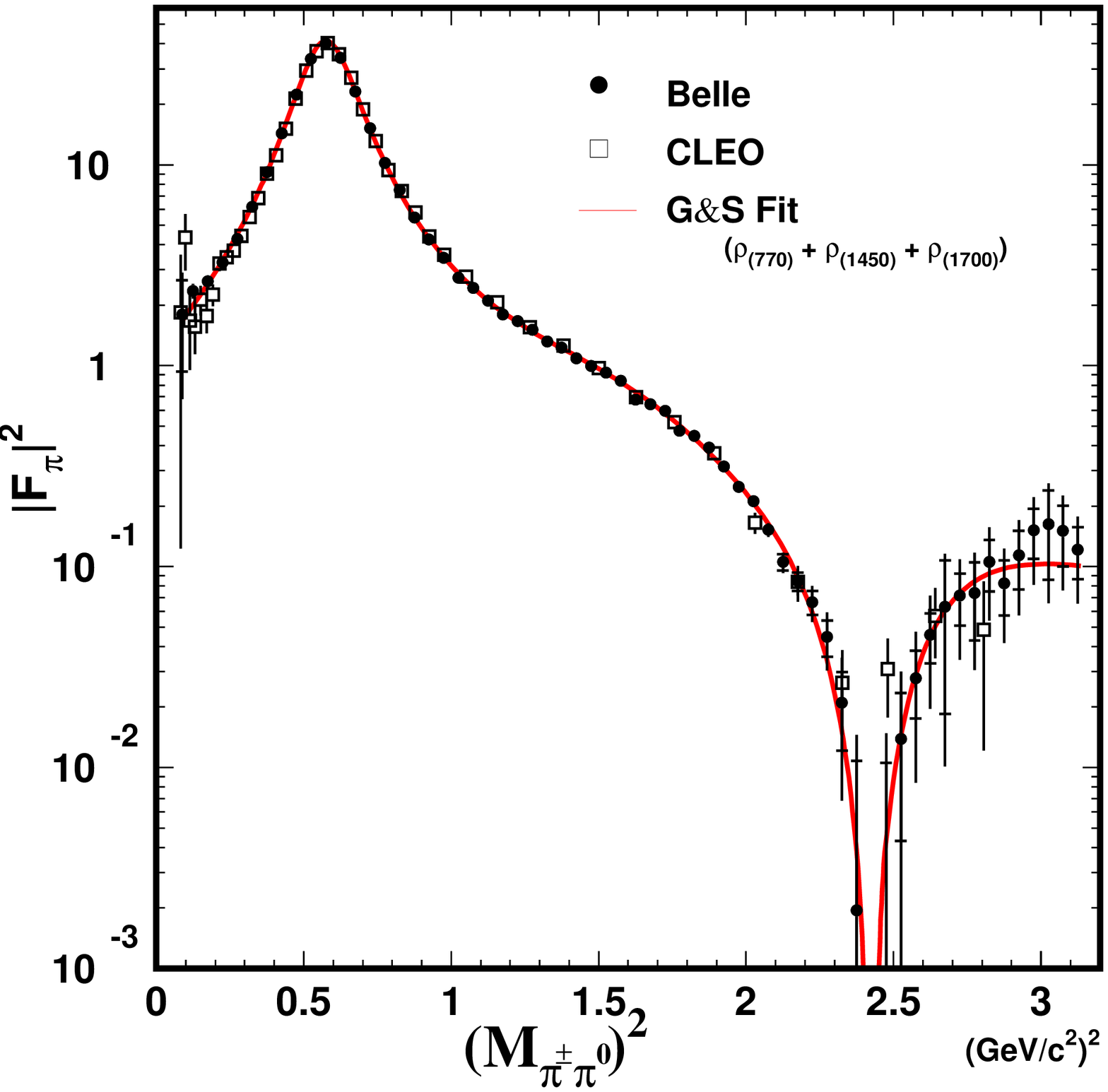}}
\caption
{ Pion form factor for $\Taupipi0$.
The solid circles show the Belle result and the
open squares show the CLEO result~\protect{\cite{CLEO2000}}.
%The error bars for Belle data include both 
%statistical and systematics added quadrature.
The error bars for the Belle data include both
statistical and systematic errors added in quadrature.
The solid curve is the 
result of a fit to the Gounaris-Sakurai model with 
the
$\rho(770)$, $\rho'(1450)$, and $\rho''(1700)$ resonances,
where all parameters are floated.
}
\label{Fpi}
\end{center}
\end{figure}

\begin{figure}[t]
\begin{center}
\rotatebox{0}{\includegraphics*[width=0.6\textwidth,clip]
%
%
%{./Fig/EPS2005fig/g5030/stat+sys2/correct_pionf3_fit_emat_g5030_set40_a.eps}}
{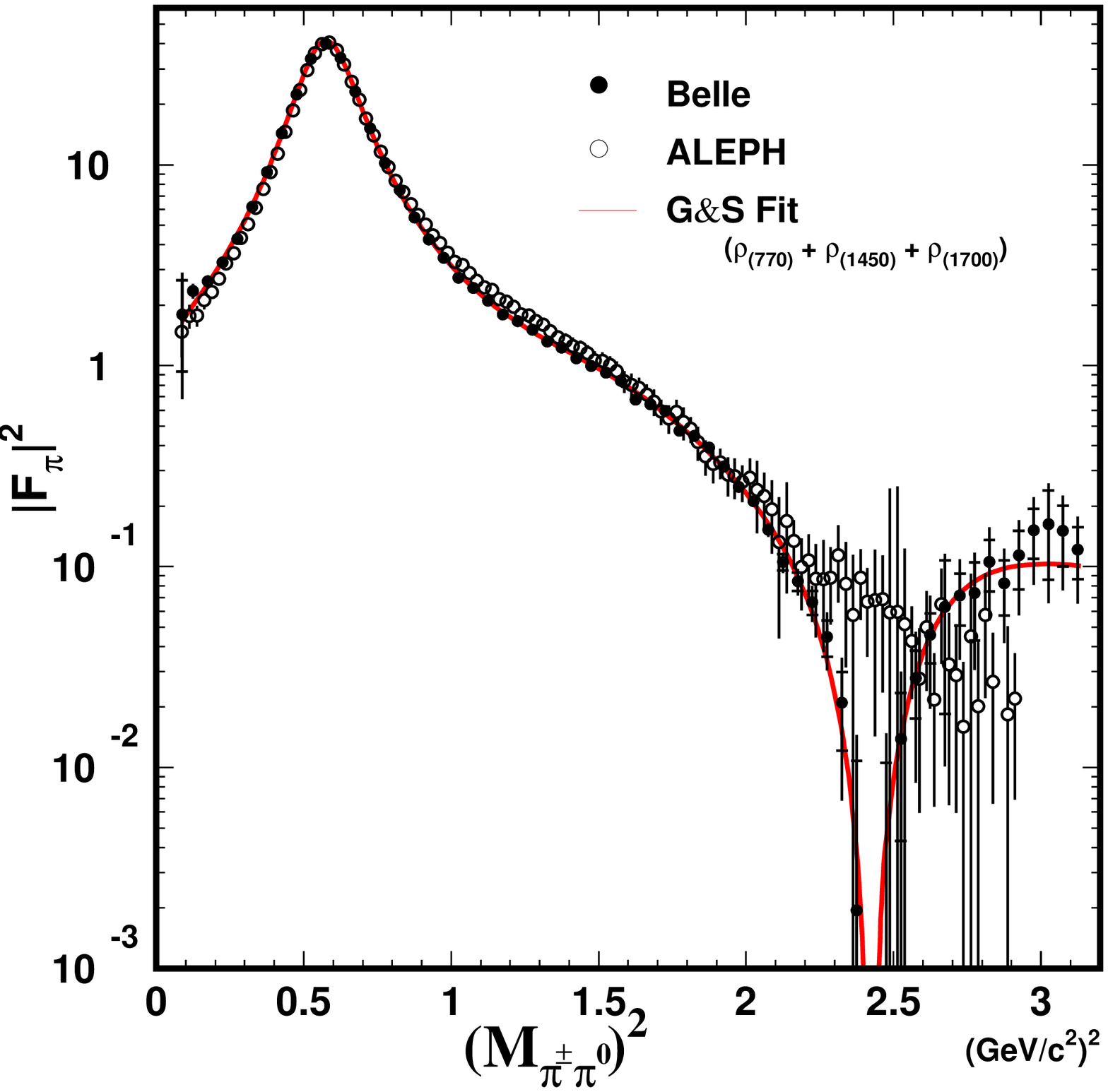}}
\caption
{ Pion form factor for $\Taupipi0$.
The solid circles show the Belle result and the
open squares show the ALEPH result~\protect{\cite{ALEPH05}}. 
%      The error bars for Belle data include both 
%      statistical and systematics added quadrature.
The error bars for the Belle data include both
statistical and systematic errors added in quadrature.
The solid curve is the 
result of a fit to the Gounaris-Sakurai model, where
all parameters are floated.
}
\label{FpiALEPH}
\end{center}
\end{figure}

\begin{figure}[t]
\begin{center}
\rotatebox{0}{\includegraphics*[width=0.6\textwidth,clip]
%
%
%{./Fig/EPS2005fig/g5030/stat+sys2/correct_pionf3_fit_res_g5030_set40_c.eps}}
{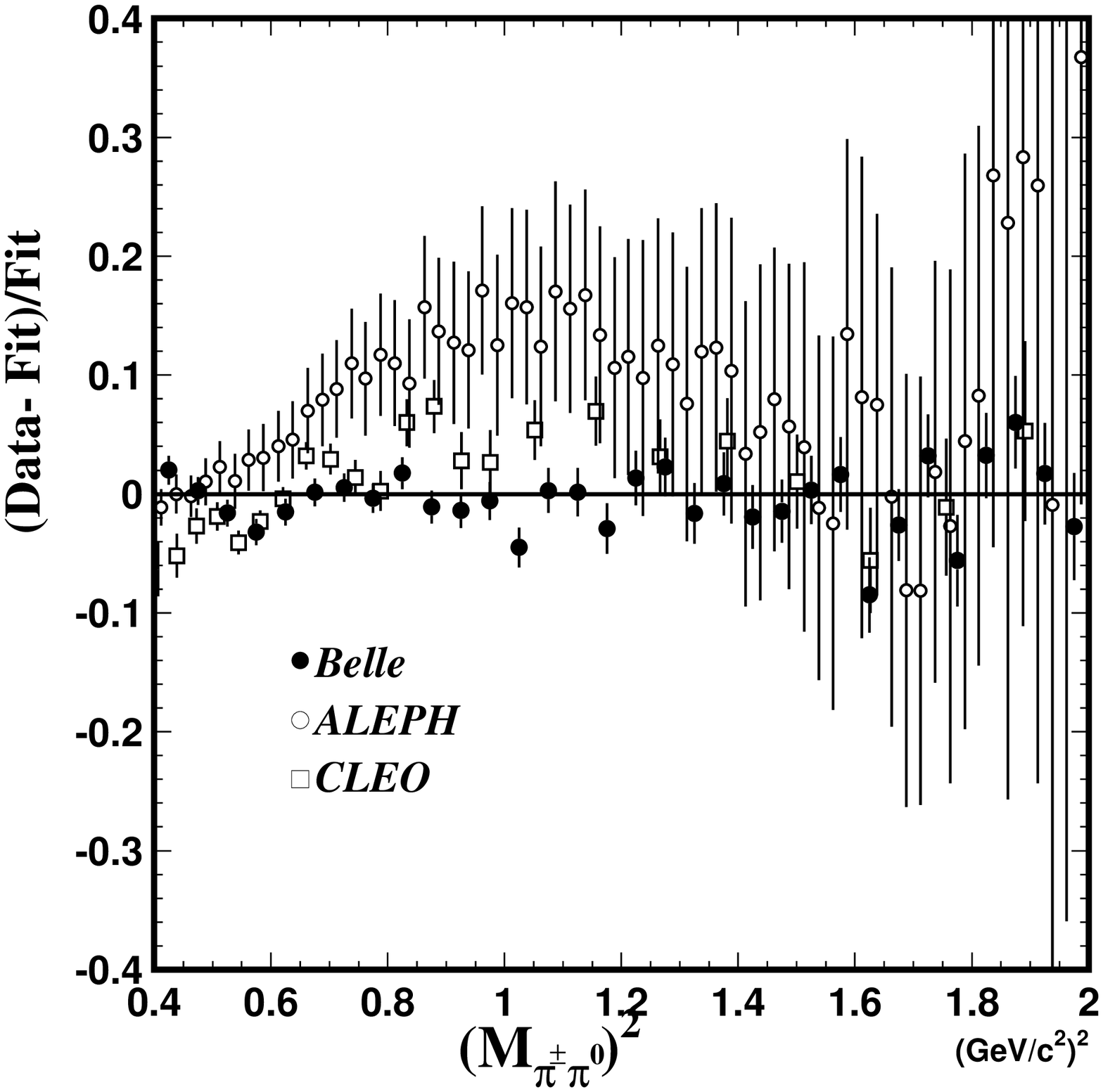}}
\caption
{ 
Comparison of the pion form factor squared $|F_{\pi}(s)|^{2}$ 
measured  
by Belle to that measured by CLEO~\protect{\cite{CLEO2000}}
 and ALEPH~\protect{\cite{ALEPH05}} 
experiments
 in the $\rho(770)$ 
and $\rho^{\prime}(1430)$ mass region.  Difference from the fit of  
 the Belle $\Taupipi0$ data divided by the fit value is plotted.
}
\label{Fpi_comp}
\end{center}
\end{figure}

The results are shown in terms of the pion form factor squared
($|F_{\pi}(s)|^{2}$) in Figs.~\ref{Fpi} and~\ref{FpiALEPH}.
A dip caused by destructive interference between the $\rho'(1450)$ and 
$\rho''(1700)$ is clearly visible.
For the first time production of the $\rho''(1700)$ 
in $\tau^-$ decays has been unambiguously demonstrated 
and its parameters determined.
For comparison, the figures also show results from the 
CLEO~\cite{CLEO2000} and ALEPH~\cite{ALEPH05} experiments, 
respectively; there is good agreement with both data sets.
%Our results show a good agreement with the CLEO and ALEPH data
% but our data are
%more precise at the $\rho''$ mass region.
%In Fig.~\ref{FpiALEPH}, our reuslts are compared with recent ALEPH data
%~\cite{ALEPH05}. Again a good agreement 
Figure \ref{Fpi_comp} shows our data and that of
CLEO for the mass range 0.2--2.2~$\GeVCC$, where the
contribution to $a^{\rm had,LO}_\mu$ 
is largest.

%------------------------- |F_pi| ----------------------------%
%%-------------------------------------------------------------------------%

%------------------------- |F_pi| ----------------------------%
%%-------------------------------------------------------------------------%

%------------------------- |F_pi| (data-fit)/fit----------------------------%
%%-------------------------------------------------------------------------%

%------------------------- |F_pi| (data-fit)/fit----------------------------%
%  \begin{figure}[t]
%   \begin{center}
%   \rotatebox{0}{\includegraphics*[width=0.6\textwidth,clip]
%  {./Fig/EPS2005fig/pionf3_fit_res_g5024_set43_a.eps}}
%   \caption
%    { Comparison of the pion form factor squared $|F_{\pi}(s)|^{2}$ 
%measured by Belle to that of recent ALEPH~\cite{ALEPH05} data}. 
% }
%
%   \label{Fpi_compALEPH}
%  \end{center}
% \end{figure}
%%-------------------------------------------------------------------------%
%============== 3-3: spectral function & pion form factor ====================%

\section{Evaluation of $a_{\mu}^{\pi\pi}$}

Using the unfolded $s= M^2_{(\pi\pi^0\,{\rm unf.})}$ distribution
$(1/N)(dN/ds)$, the spectral function $v^{-}$ is obtained
%from Eq.~(\ref{eq:specmeas}). 
by taking the inverse of Eq.~(\ref{eq:tauspec}):
\begin{eqnarray}
v_{-} & = & 
\frac {m_{\tau}^{2}} { 6\pi |V_{ud}|^{2} S_{EW} } 
\left(
\frac{ \mathcal{B}_{\pi\pi} } {\mathcal{B}_{e}} \right)
\left[     \left( 1 - \frac{s}{m_{\tau}^{2}} \right)^{2}
     \left( 1 + \frac{2s}{m_{\tau}^{2}} \right)
\right]^{-1}
\frac{1}{N}\frac{dN}{ds}\,.
\label{eq:specmeas} 
\end{eqnarray}
The resulting function can be inserted into Eq.~(\ref{eq:amu2pi})
to obtain 
the dominant low-mass contribution to
the hadronic part of the anomalous magnetic moment, 
$a_{\mu}^{\pi\pi}$. This assumes the CVC relation (\ref{eq:cvc})
holds.

There are several external parameters in these equations; 
the values used for these are listed in Table~\ref{tab:amuext}.
For $m_{\tau}$, $V_{ud}$, and $\mathcal{B}_{e}$, 
PDG~\cite{PDG2004} values are used.
For the electroweak  radiative correction $S_{\rm EW}$,
we use the recent value $1.0233\pm 0.0006$, which is 
% instead of an older value of  $S_{EW}=1.0194$.
based on a consistent treatment of 
the isospin-breaking correction~\cite{ISB2001,DEHZ}.
For the $\pi^{-}\pi^{0}$ branching fraction, 
our measurement is consistent with the world average given
in Ref.~\cite{DATAU02}. 

Including our result and the recent ALEPH $\mathcal{B}_{\pi\pi^{0}}$
measurement, the new world average is 
\begin{equation}
\mathcal{B}_{\pi\pi^{0}}= (25.42\pm 0.11)\%.
\end{equation}
We use this new world average for the evaluation of
$a_{\mu}^{\pi\pi}$.

%%(Note: can we combine the world average with
%%our measurement and use that?)
%Since our measurement of the $\pi\pi^{0}$ 
%branching fraction is consistent with the
%world average, we 
%use the most recent world average
%given in the ref.~\cite{DATAU02}.

\begin{table*}[!ht]
\begin{center}
\begin{tabular}{l|c|c|c|c}
%\multicolumn{2}{c|}{term} & value  \\ 
\hline \hline
Source & Value & Relative error   &  $\Delta a_{\mu}^{\pi\pi}$ & Reference \\
&       &       (\%)    &   $(10^{-10})$             &       \\
\hline
\hline
% Normalization Factors &            &                   \\
$S_{EW}$    &  $1.0233 \pm 0.0006$     &  0.06   &  $\pm$ 0.32   & 
 ~\cite{DEHZ},\cite{ISB2001} \\
$V_{ud}$    &  $0.9734 \pm 0.0008$      & 0.08  &   $\pm$ 0.42   & 
\cite{PDG2004}  \\
$\mathcal{B}_{e}$    & ($17.84\pm\,0.06$)\%  &  0.34 &  $\pm$ 1.82   & 
\cite{PDG2004}  \\
$\mathcal{B}_{\pi\pi^{0}}$ & ($25.42\pm\,0.11$)\% & 0.43 &$\pm$ 2.30   &
%~\cite{DATAU02}  \\
\\
\hline
Total external\ \ &                       &           & $\pm$ 3.0 &    \\
\hline
\hline
\end{tabular}
\caption {  Values of the external parameters and the 
systematic errors for $a_\mu^{\pi\pi}$ arising from 
these sources.}
\label{tab:amuext}
\end{center}
\end{table*}

%To avoid model dependence,
The integration in Eq.~(\ref{eq:amu2pi}) is carried out numerically 
by taking the sum of the integrand evaluated at the center of each bin.
%in each bin using the middle value in each bin in $s$.
The statistical error in $a_{\mu}^{\pi\pi}$ is calculated
including the off-diagonal elements of the covariance 
matrix $X_{ij}$:

\begin{eqnarray}
\delta a^{\pi\pi}_\mu & = & \sum_{i,j} 
\left( \frac{\partial a_{\mu}}{\partial\alpha_{i}} \right)
X_{ij}                             
\left( \frac{\partial a_{\mu}}{\partial\alpha_{j}} \right).
%\bigtriangleup a =  \sum_{i,j} 
%                             \left( \frac{d a_{\mu}}{d a_{i}} \right)
% X_{ij}                             
%\left( \frac{d a_{\mu}}{d a_{j}} \right).
\end{eqnarray}

Because of uncertainties associated with the background 
estimate and with the acceptance correction in the 
lowest
mass region, the integration is carried out over the 
mass range $\sqrt{s}=$ 0.50--1.80~$\GeVCC$.

\subsection{Systematic uncertainty}

Systematic uncertainty in $a_{\mu}^{\pi\pi}$ arises
from both external and internal sources. The errors
arising from external parameters are
%size of the external errors determined from the input parameters is
summarized in Table \ref{tab:amuext}; the total systematic
error from these sources is $\pm 3.0\times 10^{-10}$
(dominated by $\delta \mathcal{B}_{\pi\pi^{0}}$).

% The total systematic error from the external factors (normalization factors)
% is estimated to be $\pm 3.0\times 10^{-10}$, in which
% the error is dominated by  the $\mathcal{B}_{\pi\pi^{0}}$ measurement.

The systematic error arising from internal sources (specific 
to this measurement) are listed in Table~\ref{tab:amuerror2}.
There are two sources of background in the $\pi^{-}\pi^{0}$ sample:
(i) feed-down from  $\tau^{-}\rightarrow h^{-} (n\pi^{0})\nu_{\tau}$ and
$\tau^{-}\rightarrow K^{-}\pi^{0}\nu_{\tau}$, and
(ii) non-$\tau$ background.
In the first case, the uncertainty in the branching fraction
%
% $\bigtriangleup\mathcal{B}/\mathcal{B}(\ge 2\pi^{0}\pi)=\pm 1.5\%$,
% $\bigtriangleup\mathcal{B}/\mathcal{B}(K\pi^{0})=\pm 6.7\%$,
%
is used to estimate the error.
% $\bigtriangleup a_{\mu}=(\pm 0.3$ and $1.2)\times 10^{-10}$.
In the second case,
the uncertainty in the background as estimated from the control
samples is assigned as the error.
As mentioned earlier, the fake-$\pi^{0}$ background 
is subtracted using sideband events; the uncertainty 
is determined by varying the signal and sideband regions.

\begin{table}[!ht]
\begin{tabular}{l|c}
\hline \hline
Source    &     $\Delta a_{\mu}^{\pi\pi} \times 10^{10}$   \\
      &    (0.50--1.80~$\GeVCC$) \\
\hline\hline
Background:                      &        \\
\ \ \  non-$\tau$\,($e^+e^-\ra\bar{q}q$)   &    $\pm 0.11$    \\
\ \ \ feed-down $h(n\pi^{0})\nu$     &  $\pm 0.09$   \\
\ \ \ feed-down $K^{-}\pi^{0}\nu$       & $\pm 0.15$  \\
Energy scale                     &  $\pm 0.10 $     \\
$\pi^{0}/\gamma$ selection       &   $\pm 0.24$    \\
$\gamma$ veto                         & $\pm 0.93 $   \\
Efficiency:                      &       \\
\ \ \  $\pi^{0}/\gamma$                 &  $\pm 0.35 $     \\
\ \ \ \ charged track                   &    $ <0.10$      \\ 
Integration procedure\ \            &    $<0.10$     \\
\hline
Total internal\ \                                 & 
$\pm$ 1.04    \\
\hline\hline
\end{tabular}
\begin{center}
\caption { Systematic errors for $a_\mu^{\pi\pi}$ arising from
internal sources (specific to this measurement).}
\label{tab:amuerror2}
\end{center}
\end{table}

%\subsubsection{ Efficiency }

The ratio of the branching fractions for the decays 
$D^{0}\rightarrow K^{-}\pi^{-}\pi^{0}$ and 
$D^{0}\rightarrow K^{-}\pi^{+}$ is used to monitor 
the $\pi^{0}$ efficiency. 
%The uncertainty is estimated from the uncertainty 
%of the calibrated $\pi^{0}$ efficiency.
It is found that the shape of the mass spectrum is insensitive 
to uncertainty in the $\pi^{0}$ efficiency, as it is only
at the few \% level. 
Adding all individual errors in quadrature gives a total
error on $a^{\pi\pi}_{\mu}$ arising from 
internal sources of 
%$\bigtriangleup a_{\mu}= 
$\pm 1.0\times 10^{-10}$. 

%\subsubsection{ Other stability checks }

To check the stability of $a_{\mu}^{\pi\pi}$, we perform 
the following tests:
%In order to check that the results are sable, 
%various checks have been carried out.
\begin{enumerate}
\item 
The sample is divided into subsamples
based on the tag-side topology, i.e., 
one electron, one-prong, or three-prong.
The values of $a^{\pi\pi}_{\mu}$ obtained from these 
subsamples are consistent within the statistical errors.
\item 
The sample is divided into subsamples based on the
running period, e.g., years 2000, 2001, or 2002.
Again, the values of $a^{\pi\pi}_\mu$ obtained are 
consistent within the statistical errors.
\item The sample might be  sensitive to the 
requirement on the overlap region between
the projection of the charged track and $\gamma$ clusters. To 
estimate this sensitivity, we select events with a tighter isolation 
requirement on $\gamma$'s and on the track extrapolation: 
50~cm instead of 20~cm.
%   In order to estimate this effect,  the sample is prepared with the
%  tighter isolation requirement of the gamma and the track extrapolation.
%  The isolation condition is required to be 50 cm(default is 20 cm).
%  Once again the resultant $a_{\mu}$ agrees with the one of the
%  standard cut.

The resulting variation in $a^{\pi\pi}_{\mu}$ is small 
and is included as an additional systematic error.
\end{enumerate}

\section{Results} 

The result for $a_{\mu}^{\pi\pi}$ integrated over the mass range
%
% Br_pipi0 = 25.42 +/- 0.11 (includeing Belle and Aleph) is used
% for a_mu^{pipi}:   2005/11/04
%
%
$\sqrt{s}=$0.50--1.80~$\GeVCC$ is
\begin{eqnarray}
a_{\mu}^{\pi\pi}[0.50, 1.80] = ( 464.4 \pm 0.6\,{\rm (stat.)}
\pm 1.0\,{\rm (sys.)} \pm 3.0\,{\rm (sys.\,ext.)} \times 10^{-10},
\nonumber 
\end{eqnarray}
where the first error is statistical and the second and third errors
are systematic errors arising from internal and external sources, 
respectively.
In addition, there is a systematic uncertainty 
caused by isospin violation effects arising 
from $\rho$-$\omega$ interference, 
from the $\pi^-$ and $\pi^{0}$ mass difference, and from
radiative corrections (see~Ref.~\cite{ISB2001}). 
The overall correction is estimated to be 
$(-1.8\pm 2.3) \times 10^{-10}$,
where the central value is taken from Ref.~\cite{CLEO2000}
and we enlarged the error according to the value in Table 5 
of Ref.~\cite{DEHZ};
this correction is small because the threshold region is not included.
Applying this correction gives
\begin{eqnarray}
a_{\mu}^{\pi\pi}[0.50,1.80] = ( 462.6 
\pm 0.6\,{\rm (stat.)} \pm 3.2\,{\rm (sys.)}
  \pm 2.3\,{\rm (isospin)} ) \times 10^{-10}\,,
\nonumber 
\end{eqnarray}
where the first error is statistical, the second is 
systematic, and the third arises from isospin violation.

This result can be compared to those from previous $\tau$~\cite{DEHZ} 
and $e^+e^-$ experiments~\cite{DAV2003}:
\begin{eqnarray*}
 a_{\mu}^{\pi\pi}[0.50,1.80]& = &
 ( 464.0 \pm 3.0\,{\rm (exp.)} \pm 2.3\,{\rm (isospin)})
 \times 10^{-10} \quad (\tau:{\rm  ALEPH,CLEO}) \\
 a_{\mu}^{\pi\pi}[0.50,1.80]& = &
%%%%
% ( 450.2 \pm 4.9\,{\rm (exp.)} \pm 1.6\,{\rm (rad.)})
% \times 10^{-10} \quad (e^+e^-:{\rm CMD2,KLOE})\,.
%%%%
%%    Eur.Phys.C27,497(2003)[hep-ph/0208177 v3 ]
%
%
% ( 440.8 \pm 4.7\,{\rm (exp.)} \pm 1.5\,{\rm (rad.)})
% \times 10^{-10} \quad\quad (e^+e^-:{\rm CMD2})\,.
%%
%%    A.~Hocker, Proc. of the 32nd Int. Conf. on High Energy
%    Physics (ICHEP04), Beijing, 2004, vol2, p.710
%
 ( 448.3 \pm 4.1\,{\rm (exp.)} \pm 1.6\,{\rm (rad.)})
 \times 10^{-10} \quad\quad (e^+e^-:{\rm CMD2,KLOE})\,.
\end{eqnarray*}
The first error includes both statistical and experimental systematic errors added in  quadrature.
The second error in the $e^+e^-$ result is due
to radiative corrections.
Our result agrees well with the $\tau$-based result but is 
noticeably higher than the $e^+e^-$ result. This supports the
hypothesis that there is a difference between the mass spectra 
of the $2\pi$ systems produced in $\tau$-decay and $e^+e^-\rightarrow\pi^+\pi^-$ reactions.

%
%[DEHZ,DAV2003]   tau-based
%    energy range            a_mu(tau)
%   2m_pipi - 0.500         56.03\pm 1.61\pm 0.28
%   0.500   - 1.800        464.03\pm 3.19\pm 2.34
%  -----------------------------------------------------------
%   2m_pipi - 1.800        520.06 \pm 3.47 \pm 2.36
%                                  (\pm  4.20)
%    isospin violation correction
%     -13.8 \pm 2.4
%     is applied.
% 

\vspace*{1cm}

In summary, we have studied the decay $\Taupipi0$ using high 
statistics data
taken with the Belle detector at the KEKB $e^+e^-$ collider. 
The branching fraction is measured with 1.2\% accuracy,
which is better than that in the previous experiments
(except for the ALEPH result).
In the unfolded $\pi^-\pi^0$ mass spectrum, in addition to the 
 $\rho(770)$ and $\rho^{\prime}(1450)$ mesons, 
the production of the $\rho^{\prime\prime}(1700)$ in $\tau^{-}$ 
decays has been unambiguously demonstrated and its parameters determined. 
The unfolded spectrum is used to evaluate  
the 2$\pi$ contribution to the muon anomalous magnetic
moment $a_{\mu}^{\pi\pi}$  in the region 
$\sqrt{s}=0.50-1.80~\rm{GeV}/c^{2}$.
Our results agree well 
with the previous $\tau$ based results but 
are higher than the $e^+e^-$ results.

\section*{Acknowledgments}

We thank M. Davier and J. H. K\"{u}hn for
their advice and encouragement during this analysis. 
We thank the KEKB accelerator group for the excellent
operation of the KEKB accelerator.
We acknowledge support from the Ministry of Education,
Culture, Sports, Science, and Technology of Japan
and the Japan Society for the Promotion of Science;
the Australian Research Council
and the Australian Department of Industry, Science and Resources;
the National Science Foundation of China under contract No.~10175071;
the Department of Science and Technology of India;
the BK21 program of the Ministry of Education of Korea
and the CHEP SRC program of the Korea Science and Engineering
Foundation;
the Polish State Committee for Scientific Research
under contract No.~2P03B 01324;
the Ministry of Science and Technology of the Russian Federation;
the Ministry of Education, Science and Sport of the Republic of
Slovenia;
the National Science Council and the Ministry of Education of Taiwan;
and the U.S.\ Department of Energy.


\begin{thebibliography}{99}
\bibitem{CC}
Throughout this paper, 
the inclusion of the charge-conjugate decay mode is implied
unless stated otherwise.
\bibitem{KERNEL}
Using a variable $x$ defined as
$$
x=(1-\beta_\mu)/(1+\beta_\mu), \quad\quad  
\beta_\mu=(1-4m_\mu^2/s)^{\frac{1}{2}}\,, \nonumber
$$
$\hat{K}(s)$ is given by
\begin{eqnarray}
 \hat{K}(s)&=& \frac{3s}{m_{\mu}^{2}} \cdot K(s) \nonumber \\
 K(s)&=& x^2\left(1-\frac{x^2}{2}\right)+(1+x)^2\left(1+\frac{1}{x^2}\right)
   \nonumber \\
 && \left(\ln( 1 + x )  - x   
+ \frac{x^2}{2}\right)
+\left( \frac{1+x}{1-x}\right)x^2\ln x. \nonumber
%\label{eqn:kfunc}
\end{eqnarray}
\end{thebibliography}
\end{document}